\documentclass[12pt,letterpaper,]{article}
\usepackage{jheppub}

\usepackage{graphics,epsfig}
\usepackage{graphicx, color}
\usepackage{slashed}
\usepackage{amsmath}
\usepackage{amssymb}
\usepackage{bbm}
\usepackage{epsfig}
\usepackage{yfonts}
\usepackage{graphicx}
\usepackage{caption}
\usepackage{subcaption}
\newcommand{\labell}[1]{\label{#1}}

\newcommand{\Li}{\textrm{Li}}

\newcommand{\be}{\begin{equation}}
\newcommand{\ee}{\end{equation}}
\newcommand{\bea}{\begin{eqnarray}}
\newcommand{\eea}{\end{eqnarray}}
\newcommand{\ba}{\begin{eqnarray}}
\newcommand{\ea}{\end{eqnarray}}

\newcommand{\beq}{\begin{equation}}
\newcommand{\eeq}{\end{equation}}
\newcommand{\beqa}{\begin{eqnarray}}
\newcommand{\eeqa}{\end{eqnarray}}
\newcommand{\beqar}{\begin{eqnarray*}}
\newcommand{\eeqar}{\end{eqnarray*}}

\newcommand{\reef}[1]{(\ref{#1})}

\newcommand{\eg}{{\it e.g.,}\ }
\newcommand{\ie}{{\it i.e.,}\ }

\newcommand{\E}{\mathcal{E}}
\newcommand{\R}{\mathcal{R}}
\newcommand{\N}{\mathcal{N}}

\newcommand{\lp}{\ell_{\mt P}}

\newcommand\te{t_\mt{E}}

\newcommand{\ren}{R\'enyi\ }

\newcommand{\tr}{{\rm tr}}
\newcommand{\Tr}{{\rm Tr}}

\def\({\left(} \def\){\right)}
\def\[{\left[} \def\]{\right]}

\def\pd{\partial}
\def\=d{\, {\buildrel \rm def  \over =} \,}

\def\sqr#1#2{{\vcenter{\vbox{\hrule height.#2pt \hbox{\vrule width.#2pt height#1pt \kern#1pt \vrule width.#2pt}\hrule height.#2pt}}}}

\def\beq#1{\begin{equation} \label{#1}}
\def\ben{\begin{equation*}}
\def\een{\end{equation*}}
\def\bequa{\begin{eqnarray}}
\def\eequa{\end{eqnarray}}

\def\Tr{\mathop{\mathrm{Tr}}}

\def\beq{\begin{equation}}

\def\al{\alpha}

\newcommand{\bseq}{\begin{subequations}}
\newcommand{\eseq}{\end{subequations}}

\renewcommand{\tanh}{\mathop{\rm th}\nolimits}

\renewcommand{\ln}{\mathop{\rm ln}\nolimits}

\newcommand{\mt}[1]{\textrm{\tiny #1}}
\newcommand{\comment}[1]{\textcolor{red}{\bf [[[#1]]]}}

\newcommand{\rhoa}{\rho_\mt{A}}
\newcommand{\mue}{\mu}  
\newcommand{\imu}{\mu_\mt{E}} 
\newcommand{\iq}{q_\mt{E}}
\newcommand{\norm}{{n}_\mt{A}}
\newcommand{\tS}{\tilde{S}}
\newcommand{\tnorm}{{\tilde n}_\mt{A}}
\newcommand{\taue}{\tau_\mt{E}}
\newcommand{\tsigma}{\tilde{\sigma}}
\newcommand{\lstar}{\ell_*}

\newcommand{\ha}{\hat{a}}
\newcommand{\hb}{\hat{b}}
\newcommand{\hc}{\hat{c}}
\newcommand{\he}{\hat{e}}

\arxivnumber{1310.4180}

\title{Holographic Charged R\'enyi Entropies}

\author[a,b]{Alexandre Belin,}
\author[b]{Ling-Yan Hung,}
\author[a,b]{Alexander Maloney,}
\author[a,b]{Shunji Matsuura,}
\author[c]{Robert C. Myers}
\author[c,d]{and Todd Sierens}

\affiliation[a]{Departments of Physics and Mathematics, McGill
University, Montr\'eal, Qu\'ebec, Canada}
\affiliation[b]{Department of Physics, Harvard University,
Cambridge, MA 02138 USA}
\affiliation[c]{Perimeter Institute for Theoretical Physics,
Waterloo, Ontario N2L 2Y5, Canada}
\affiliation[d]{Department of Physics \& Astronomy and
Guelph-Waterloo Physics Institute,
University of Waterloo, Waterloo, Ontario N2L 3G1, Canada}

\emailAdd{alexandre.belin@mail.mcgill.ca}
\emailAdd{lhung@physics.harvard.edu}
\emailAdd{maloney@physics.mcgill.ca}
\emailAdd{matsuura@physics.mcgill.ca}
\emailAdd{rmyers@perimeterinstitute.ca}
\emailAdd{tsierens@perimeterinstitute.ca}

\abstract{We construct a new class of entanglement measures by extending the usual
definition of \ren entropy to include a chemical potential. These charged \ren
entropies measure the degree of entanglement in different charge sectors of the
theory and are given by Euclidean path integrals with the insertion of a Wilson
line encircling the entangling surface. We compute these entropies for a
spherical entangling surface in CFT's with holographic duals, where they are
related to entropies of charged black holes with hyperbolic horizons. We also
compute charged \ren entropies in free field theories.}

\begin{document}

\maketitle


\section{Introduction}

Entanglement and \ren
 entropies have emerged as diagnostic probes of
considerable practical and formal interest in areas ranging from condensed
matter physics, \eg \cite{wenx,cardy0} to quantum gravity, \eg
\cite{rt0,mvr,arch}. In this paper we will consider a generalization of these
entropies for systems with a conserved global charge.

Consider a quantum system consisting of two components, A and B,  in a state
described by the density matrix $\rho$. We will consider quantum field
theories, where A and B are spatial regions separated by an entangling surface
$\Sigma$. We then trace over the degrees of freedom in region B to construct
the reduced density matrix $\rho_\mt{A} = \Tr_{\rm B}\rho$. The latter contains
information about the entanglement between A and B. The \ren entropies
\cite{renyi0,renyi1}
 \be
S_n = {1\over 1-n}\,\log\, \Tr \rho_\mt{A}^{\,n}\,,
 \labell{ren0}
 \ee
are the moments of this reduced density matrix. The limit $n\to1$ gives the
entanglement entropy, $S_\mt{EE}=\lim_{n\to1} S_n =-\Tr\!\[ \rho_\mt{A} \log
\rho_\mt{A}\]$.

In this paper, we will consider quantum field theories with a
conserved (global) charge. In this case we can ask whether the entanglement
between A and B depends on how charge is distributed between the two
subsystems.  This is characterized by the following `grand canonical'
generalization of the \ren entropy:
 \ba
S_{n}(\mue) = {1\over 1-n}\log \Tr \left[\,\rhoa\, \frac{e^{\mue\,
Q_\mt{A}}}{\norm(\mue)}\right]^n\,.
 \labell{charen}
 \eea
Here $\mue$ is a chemical potential conjugate to $Q_\mt{A}$, the charge
contained in subsystem A. We have also introduced $\norm(\mue)\equiv \Tr\left[
\,\rhoa\, e^{\mue\, Q_\mt{A}}\right]$ to ensure that the new density matrix
(enclosed by the square brackets above) is properly normalized with unit trace.
The $\mue$-dependence of these `charged' \ren entropies $S_{n}(\mue)$
encodes the dependence of the entanglement on the charge.

We will also be interested in the entropies constructed
with an imaginary chemical potential
 \ba
\tS_n(\imu)={1\over 1-n}\log \Tr\left[\,\rhoa\, \frac{e^{i\imu\,
Q_\mt{A}}}{\tnorm(\imu)}\right]^n\,, \labell{charen2}
 \eea
where $\imu$ is real and  $\tnorm(\mue)\equiv \Tr\left[ \,\rhoa\, e^{i\imu\,
Q_\mt{A}}\right]$. As we will see below, the analytic continuation between
\eqref{charen} and \eqref{charen2}
is typically
straightforward in the vicinity of the origin $\mu=0$, but one typically encounters an
interesting structure of singularities along the imaginary $\mue$-axis.

In quantum field theory, \ren entropies can be evaluated using the replica
trick \cite{cardy0}, which relates them to a Euclidean path integral on an
$n$-sheeted geometry. These path integral calculations of $S_n$ are easily
extended to compute our new charged \ren entropies. As we will describe below,
the key new ingredient  is a Wilson line encircling the entangling surface.
This generalizes the twist operator $\sigma_n$ appearing in the replica trick
to include a `magnetic flux' proportional to $\mu$.

In conformal field theories, these entropies can be studied rather explicitly.
A useful approach was introduced in \cite{casini9} to evaluate the entanglement
entropy across a spherical entangling surface for an arbitrary $d$-dimensional
CFT in flat space. The latter entropy is related by a conformal mapping to the
thermal entropy of the CFT on a hyperbolic cylinder $R\times H^{d-1}$, where
the temperature and curvature are fixed by the radius of the original
entangling surface. If the temperature is allowed to vary, the thermal entropy
calculates \ren entropies \cite{renyi,subir}. In theories with holographic
gravity duals, the thermal entropy is the horizon entropy of a
black hole with  $H^{d-1}$ horizon. Our charged entropies $S_{n}(\mue)$ can
also be computed for spherical entangling surfaces with a simple extension of
this procedure. The same conformal
mapping leads to a grand canonical ensemble with chemical potential $\mue$ for
the CFT on the hyperbolic cylinder. In the holographic context, the presence of
a global symmetry in the boundary CFT leads to a gauge field in the dual
gravity theory. $S_{n}(\mue)$ is then related to the entropy of a hyperbolic
black hole which is charged under this gauge field.

This paper is organized as follows: In section \ref{two}, we will discuss
general features of the charged \ren entropy and outline its computation in
CFT.  We also describe the various properties of the corresponding twist
operators. In section \ref{Einstein}, we compute $S_{n}(\mue)$ in holographic
CFT's (in spacetime dimensions $d\ge3$) by considering the charged hyperbolic
black holes of the dual Einstein-Maxwell theory. We conclude with a discussion
and general comments on the properties of the charge \ren entropy in section
\ref{discuss}. Three appendices are also included. Appendix \ref{free}
describes computations of $\tS_{n}(\imu)$ in simple free field theories and
appendix \ref{threed} gives the holographic computation in $AdS{}_3/CFT_2$. The
latter case is notable in that the dependence of the charged \ren entropy on
$n$ agrees with the free field theory result. Finally appendix \ref{details}
contains various details of the holographic calculations, which are used in
section \ref{Einstein}.

Charged \ren entropies arose in several recent papers which appeared while this
paper was in preparation. First, they were briefly considered in
\cite{juan,diana}. Charged \ren entropies appear in \cite{gm}, which
investigates the dynamical evolution of entanglement entropy in two-dimensional
CFT's. The supersymmetric \ren entropies three-dimensional ${\cal N}\ge2$
superconformal theores calculated in \cite{susy} can be cast in the form
\reef{charen2} with an imaginary chemical potential for a circular entangling
surface, using the conformal mappings presented in \cite{casini9,renyi}. In
this case, $Q_\mt{A}$ corresponds to the $R$-charge of the underlying theory.

\section{Charged \ren entropies for CFT's} \labell{two}

In this section, we begin with some general comments about the charged \ren
entropies \reef{charen}.  We then focus on their computation in conformal field
theories, extending the approach of \cite{casini9,renyi}, to relate charged
\ren entropies for spherical entangling surfaces in flat space to the thermal
entropy of the CFT on $R\times H^{d-1}$ in grand canonical ensemble. We also
describe the calculation of the conformal weight and magnetic response of the
corresponding twist operator. Finally, we describe the computation of these
entropies for two-dimensional CFT using twist operators.

\subsection{Replica Trick}
To begin, let us recall the replica trick \cite{cardy0,rt0}. For simplicity,
we focus on entanglements in the ground state of a QFT in $d$-dimensional flat
space. One begins by introducing an entangling surface $\Sigma$ which divides
the spatial slice at time $\te=0$ into two regions, A and
B.\footnote{Implicitly, an initial step in these calculations is to Wick rotate
the time coordinate: $\te=it$.} Integer powers of the reduced density matrix
$\rhoa$ are represented by a Euclidean path integral
 \be
\Tr\,\rhoa^n=Z_n/(Z_1)^n\,.
 \labell{replica1}
 \ee
Here $Z_n$ is the partition function on an $n$-fold cover of (Euclidean) flat
space with cuts introduced on region A at $\te=0$. At the cut, copy $n$ is
connected to copy $n$+1 when approaching from $\te\to0^-$ and to copy $n$--1
when approaching from $\te\to0^+$. In this construction, the entangling surface
$\Sigma$ becomes the branch-point of the branch-cut which separates different copies
in the n-fold covering geometry. It is convenient to think of these
boundary conditions as produced by the insertion of a ($d$--2)-dimensional
surface operator at $\Sigma$, \ie a twist operator $\sigma_n$
\cite{cardy0}.\footnote{For further discussion of twist operators beyond $d=2$,
see also \cite{renyi,tweak,twistop}.} The factors of $Z_1$ appear above to
ensure that the density matrix is properly normalized with
$Tr\left[\rhoa\right]=1$. With the usual definition \reef{ren0}, the \ren
entropies  become
  \be
S_n= {1\over n-1}\,\(n\,\log Z_1 - \log Z_n\)\,.
  \labell{ren2}
  \ee
Of course, one would  like to consider an analytic continuation to real values
of $n$ to determine the entanglement entropy with $S_\mt{EE}=\lim_{n\to1} S_n$.
Similarly, two other interesting limits are given by $n\to0$ and $\infty$,
which yield expressions which are known as the Hartley entropy and the
min-entropy, respectively \cite{head}. In particular, one finds $S_0
 = \log ({ \bf d})$ where ${\bf d}$ is the number
of nonvanishing eigenvalues of $\rho_\mt{A}$ and $S_\infty  = - \log
(\lambda_1)$ where $\lambda_1$ is the largest eigenvalue of $\rhoa$.

We wish to extend these path integral calculations of $S_n$ to compute charged \ren entropies. Let us first recall how a chemical potential
$\mu$ is included in the standard Euclidean path integral representation of a
(grand canonical) thermal ensemble. In this framework, the chemical potential
is represented by a fixed background gauge potential\footnote{We use $B_\mu$ to
distinguish this nondynamical gauge field from the bulk gauge potential $A_\mu$
appearing in our holographic calculations. This background
gauge field is imaginary in the Euclidean path integral, corresponding to a
real chemical potential.} $B_\mu$ which couples to the relevant conserved
current. Of course, in the `thermal' path integral, the Euclidean time
direction is compactified with period $\Delta\te=1/T$ and then the chemical
potential appears as a nontrivial Wilson line on this thermal circle, \ie
$\oint B = -i \mu/T$.

To evaluate the charged \ren entropies \reef{charen}, we must
compute a grand canonical version of eq.~\reef{replica1}, \ie
 \be
\Tr \left[\,\rhoa\, \frac{e^{\mue\,
Q_\mt{A}}}{\norm(\mue)}\right]^n=\frac{Z_n(\mue)}{(Z_1(\mue))^n}\,.
 \labell{replica2}
 \ee
with $\norm(\mue)\equiv \Tr\left[ \,\rhoa\, e^{\mue\, Q_\mt{A}}\right]$. Here
$Z_n(\mue)$ is computed as above, except with the insertion of a Wilson line
encircling the entangling surface $\Sigma$. That is, we  introduce a fixed
background gauge field coupling to the conserved current. This
background field is such that loops encircling the entangling surface carry a
nontrivial Wilson line, $\oint_{\cal C} B = -i n \mue$.\footnote{The
orientation of the contour will become evident in our examples below.} Here the
factor $n$ is analogous to the $1/T$ factor appearing for the thermal ensemble
above and arises here because the loop $\cal C$ circles $n$ times around
$\Sigma$, passing through all $n$ sheets of the covering geometry. The Wilson
line should be the same on all such curves and so the background gauge field is
flat, \ie $dB=0$, away from the entangling surface. By Stokes' theorem these
loops enclose a fixed flux, \ie $\int_M dB=\oint_{{\cal C}=\partial M} B = -i n
\mue$ for any two-dimensional surface pierced by $\Sigma$. Thus the entangling
surface carries a `magnetic flux' $-in\mue$ of the background gauge field. An
alternative perspective is that eq.~\reef{replica2} defines a generalized class
of twist operators $\tsigma_n(\mue)$, which are constructed by binding to the
original twist operators $\sigma_n$, a ($d$--2)-dimensional `Dirac sheet'
carrying the magnetic flux $-in\mue$. Another noteworthy comment is that with
the above definitions, the chemical potential in our charged \ren entropy
\reef{replica2} is dimensionless, in contrast to the standard chemical
potential in a thermal context, which carries the units of energy.\footnote{In
the thermal context, these units have a natural meaning by comparing the
chemical potential to the temperature. In the entanglement context, there is no
such natural reference scale with which to compare $\mue$.} In any event, given
the path integral construction describing eq.~\reef{replica2}, the charged \ren
entropies \reef{charen} become
  \be
S_{n}(\mue)= {1\over n-1}\,\(n\,\log Z_1(\mue) - \log Z_n(\mue)\)\,.
  \labell{ren3}
  \ee

Above, we considered a real chemical potential $\mue$, as would appear in
standard thermodyanmics. We will also consider analytic continuations of the
chemical potential to imaginary values, as in eq.~\reef{charen2}. As
motivation, we note that working with an imaginary chemical potential has
proven to be a useful way to probe the confinement phase transition in QCD
\cite{Roberge} and to avoid the sign problem in the lattice fermion algorithms
\cite{lattice}. Computations of the Witten index can also be interpreted in a
similar fashion \cite{Witten-Index}. Replacing $\mue=i\imu$ in our analysis
above, the Wilson loop of the background gauge field now becomes real with
$\oint_{\cal C} B = n \imu$. Further the corresponding magnetic flux carried by
the generalized twist operators is also real. Hence the effect of the imaginary
chemical potential is to introduce a simple phase as a charged operator circles
around the entangling surface. Note that the analytic continuation between real
and imaginary values requires care because, as we will see below, the partition
function has an interesting singularity structure in the complex $\mue$-plane.
We will consider both real and imaginary chemical potentials in the following.
These two cases will always be distinguished by the notation $\mue$ and $\imu$,
respectively.

\subsection{Spherical entangling surfaces} \labell{sphereX}

For the remainder of this section we will focus on computations of charged \ren
entropy in $d$-dimensional conformal field theories.  We will consider a CFT in
flat space in its vacuum state, and choose the entangling surface to be a
sphere of radius $R$ (in a constant time slice). In this case, the argument of
\cite{casini9} implies that the usual entanglement entropy equals the thermal
entropy of the CFT on a hyperbolic cylinder $R\times H^{d-1}$, where the
temperature and curvature are fixed by the radius of the original entangling
surface. A simple extension of this approach also allows one to calculate \ren
entropies \cite{renyi}. We will compute charged \ren entropies by further
extending the procedure to include a background gauge field. For simplicity of
notation we will use an imaginary chemical potential, but of course the same
construction applies for a real chemical potential.

We begin with a brief review of \cite{casini9,renyi}. First, we write the metric on
flat Euclidean space in terms of a  complex coordinate $\omega=r+i
\te$:
 \be
ds^2_{R^d} = d\omega d\bar{\omega} + \left(\frac{\omega +\bar{\omega}}{2}
\right)^2 d\Omega_{d-2}^2\,,
 \labell{flat0}
 \ee
where $\te$ is the Euclidean time coordinate, $r$ is the radial coordinate on
the constant time slices and $d\Omega_{d-2}$ is a standard round metric on a
unit $(d-2)$-sphere. The entangling surface is the sphere  at
$(\te,r)=(0,R)$.

We will perform a conformal
transformation of the above $R^d$ geometry to $H^{d-1}\times S^1$ as follows.
Introducing a second complex coordinate $\sigma = u + i \frac{\taue}{R}$, we
perform the coordinate transformation, which,
in terms of the complex coordinate $\sigma$ defined by
 \be
e^{-\sigma} = \frac{R-\omega}{R+\omega} \,. \labell{change}
 \ee
the metric \reef{flat0} then takes the form
 \be
ds^2_{R^d}= \Omega^{-2}\,R^2\left[d\sigma d\bar{\sigma} +
\sinh^2\left(\frac{\sigma + \bar{\sigma}}{2}\right) d\Omega_{d-2}^2\right]\,,
 \labell{golf}
 \ee
where
 \be
\Omega = \frac{2R^2}{|R^2-\omega^2|}=|1+\cosh\sigma|\, .
 \labell{factx}
 \ee
 The $\Omega^{-2}$ prefactor can now be removed by a simple Weyl rescaling.
 Letting $\sigma = u + i \frac{\taue}{R}$, the resulting
conformally transformed metric is
 \be
ds^2_{H^{d-1}\times S^1} = \Omega^2\, ds^2_{R^d} = d\taue^2+R^2\left(du^2 +
\sinh^2\! u\, d\Omega_{d-2}^2\right)\,. \labell{frog}
 \ee
This is  $S^1\times H^{d-1}$; $u$ is the (dimensionless) radial coordinate on the hyperboloid
$H^{d-1}$ and $\taue$ is the Euclidean time coordinate on $S^1$. The curvature radius of $H^{d-1}$ is $R$,
the radius of the original spherical entangling surface. The periodicity of the $\taue$ circle is $2\pi R$.
Note that the original entangling surface has
been pushed out to the asymptotic boundary, \ie $u\to\infty$, in the
conformally transformed geometry \reef{frog}.

The key point is that, under this conformal mapping, the density matrix describing the CFT vacuum state on the interior of the
entangling surface is transformed to a thermal
density matrix with temperature
 \beq
 T_0=\frac{1}{2\pi R}
 \labell{bus}
 \eeq
on the new hyperbolic geometry. That is, the reduced density matrix related to
the thermal density matrix as
 \beq
\rhoa=U^{-1}\, \frac{e^{-H/T_0}}{Z(T_0)}\,U \,, \labell{triangle}
 \eeq
where $U$ is the unitary transformation implementing the conformal
transformation. Since the entropy is insensitive to unitary transformations,
the desired entanglement entropy just equals the thermal
entropy in the transformed space. This same conformal mapping can also be used
to evaluate the \ren entropy. The only difference is that it would be applied
to the $n$-fold cover of flat space used to evaluate
eq.~\reef{replica1}.  In this case the period of $\taue$ is $2\pi R
n$, so the corresponding thermal ensemble has a temperature $T=T_0/n$.

We can compute charged \ren entropies \reef{charen2} by generalizing this
approach. First after having identified the appropriate charge $Q$, we
introduce a (dimensionless) chemical potential $\mue$ and the previous density
matrix \eqref{triangle} becomes
 \ba
\rho_{\text{therm}} =\frac{e^{-H/T_0+\mue Q}}{Z(T_0,\mue)} \,.
 \labell{eq:rho0x}
 \eea
Now, in fact, our discussion is slightly simplified if we consider instead an
imaginary chemical potential $\imu=-i\mue$ with which the above expression
turns into
 \ba
\rho_{\text{therm}} =\frac{e^{-H/T_0+i\imu Q}}{Z(T_0,\imu)} \,.
 \labell{eq:rho}
 \eea
As discussed above, this chemical potential is incorporated into the thermal
path integral via a background gauge field with a nontrivial Wilson line on the
Euclidean time circle:
 \beq
\imu=\oint B = \int_0^{2\pi R} B_{\taue} d\taue\,.
 \labell{round}
 \eeq
In this case the potential is just constant: $B_{\taue}=\imu/(2\pi R)$. The
background gauge field is invariant under the conformal transformation mapping
between the hyperbolic geometry and flat space. Therefore in the flat space
coordinates, we may express this gauge field as
 \beqa
B&=& \frac{i R}{2\pi}\imu\left[\frac{d\omega}{R^2-\omega^2}-
\frac{d\bar\omega}{R^2-\bar\omega^2}\right]
 \nonumber\\
 &=&-\frac{R}{\pi}\imu\,\frac{2\te
 r\,dr+(R^2 -r^2+ \te^2)\,d\te}{(R^2-r^2+\te^2)^2 +4\te^2 r^2}\,.
 \labell{gag}
 \eeqa
Now one can readily verify that this background gauge field yields $\oint
B=\imu$ for any contour encircling the entangling surface at $(\te,r)=(0,R)$ in
the flat space geometry.\footnote{An interesting exercise to gain better
intuition for this background gauge field \reef{gag} is to expand the
coordinates near the spherical entangling surface: $\te=\rho\sin\theta$ and
$r=R+\rho\cos\theta$ with $\rho\ll R$. To leading order in $\rho/R$, one then
finds that the potential reduces to $B\simeq \frac{\imu}{2\pi}d\theta$.} Of
course, we also have $dB=0$ and so this is precisely the background required to
evaluate the charged \ren entropy \reef{charen2} in this particular case.

Hence we must simply supplement the conformal mapping approach of
\cite{casini9,renyi} with the background gauge field \reef{gag} to evaluate the
charged \ren entropy across a spherical entangling surface for the CFT vacuum
in flat space. The effective reduced density matrix in eq.~\reef{charen2} is
again simply related to the thermal density matrix \reef{eq:rho} on the
hyperbolic space:
 \beq
\rhoa\, \frac{e^{i\imu\, Q_\mt{A}}}{\tnorm(\imu)}\ =\
U^{-1}\,\rho_{\text{therm}} \,U\ =\ U^{-1}\, \frac{e^{-H/T_0+i\imu
Q}}{Z(T_0,\imu)}\,U \,,
 \labell{square}
 \eeq
where as in eq.~\reef{triangle}, $U$ is the unitary transformation implementing
the conformal transformation between the two geometries. Given this expression,
the charged \ren entropy \reef{charen2} can be evaluated in terms of the
corresponding thermal partition function for the CFT on the hyperbolic geometry
as
 \beq
\tS_{n}(\imu) = {1\over 1-n}\log \frac{Z(T_0/n,\imu)}{ Z(T_0,\imu)^n}\,.
 \labell{cow}
 \eeq
Recall that in evaluating $Z(T_0/n,\imu)$, the period of $\taue$ is extended to
$2\pi Rn$, however, the gauge potential remains fixed as $B_{\taue}=\imu/(2\pi
R)$. Hence the total Wilson line around the thermal circle increases by a
factor of $n$ as desired, \ie $\oint B=n\imu$. Now using the standard
thermodynamic identity for the grand canonical ensemble
 \beq
S_{\text{therm}}(T,\imu)=-\left.{\pd F(T,\imu)\over \pd T}\right|_{\imu} ={\pd
\over \pd T}(T\log Z(T,\imu))\big|_{\imu}\,,
 \label{thermal S}
 \eeq
one easily derives the following relation between the charged \ren entropy and
the thermal entropy \cite{renyi}:
 \beq
\tS_{n}(\imu)={n\over n-1}{1\over
T_{0}}\int_{T_{0}/{n}}^{T_{0}}S_{\text{therm}}(T,\imu)\,dT\,.
 \labell{Rnyi-therm}
 \eeq

At this point, we may remind the reader that the above discussion makes no
reference to the AdS/CFT correspondence. In the special case of a holographic
CFT, this analysis may be further extended by evaluating the thermal entropy on
the hyperbolic background in terms of the horizon entropy of a topological
black hole in the bulk with a hyperbolic horizon \cite{casini9,renyi} --- see
also \cite{cthem}. In the context of the charged \ren entropy \reef{charen},
the boundary CFT also contains a conserved current corresponding to the charge
probed by these entropies and hence the bulk theory will also include a dual
gauge field. The holographic representation of the grand canonical ensemble
considered above will then be a topological black hole which is charged under
this gauge field. We will turn to such holographic calculations in section
\ref{Einstein}. First, however, we continue below with some further remarks
which apply to general CFT's.

\subsection{Properties of generalized twist operators}
\labell{PGtwist}

As discussed at the beginning of this section, the calculation of (either
ordinary or charged) \ren entropies can be viewed as involving the insertion of
a twist operator at the entangling surface. A generalized notion of conformal
dimension can be defined for these surface operators by considering the leading
singularity in the correlator $\langle T_{\mu\nu}\, \sigma_n \rangle$. This
leading singularity is fixed by symmetry, as well as the tracelessness and
conservation of the stress tensor. To be precise, consider inserting the stress
tensor $T_{\mu\nu}$ at a perpendicular distance $y$ from the twist operator
$\sigma_n$, such that $y$ is much smaller than any scales defining the geometry
of the entangling surface $\Sigma$. Then the leading singularity takes the
following form\footnote{These correlators \reef{generalT} should be normalized
by dividing by $\langle \sigma_n \rangle$. However, we leave this implicit to
avoid further clutter.}
 \ba
\langle T_{ab}\, \sigma_n \rangle &=& -\frac{h_n}{2\pi}
\frac{\delta_{ab}}{y^d}\,, \qquad \langle T_{ai}\, \sigma_n\rangle = 0\,,
\labell{generalT}\\
\langle T_{ij}\, \sigma_n\rangle &=& \frac{h_n}{2\pi} \frac{(d-1)\delta_{ij} -
d n_i n_j}{y^d}\,,
 \nonumber
 \eea
where $a,b$ ($i,j$) denote tangential (normal) directions to the twist operator
and $n_i$ is the unit vector directed orthogonally from the twist operator to
the $T_{\mu\nu}$ insertion. Thus the singularity is completely fixed up to the
constant  $h_n$, which is referred to as the conformal dimension of $\sigma_n$.
The approach reviewed in the previous section can be applied to determine the
value of $h_n$ in terms of the thermal energy density $\mathcal{E}(T,\mue)$ on
the hyperbolic cylinder \cite{renyi},\footnote{Note that in the Euclidean
background, $ \mathcal{E}(T,\mue)=-\langle T_{\taue \taue}\rangle$. Also
observe that we phrase the discussion in this subsection in terms of a real
chemical potential $\mue$.}
 \be
h_n(\mue) = \frac{2\pi n }{d-1}\,R^d\,\Big(\mathcal{E}(T_0,\mue=0)
-\mathcal{E}(T_0/n,\mue)\Big)\,.\labell{weight}
 \ee
The first term above arises because of the anomalous behaviour of the stress
tensor under conformal transformations. Of course, as is implicit above, these
remarks apply equally well for the original twist operators $\sigma_n$ and for
the generalized twist operators $\tsigma_n(\mue)$ appearing in the calculation
of the charged \ren entropy. In particular, the arguments of \cite{renyi}
yielding eq.~\reef{weight} apply without any change in the presence of the
background gauge potential.

In the context of the charged \ren entropies, another operator in the
underlying CFT is the current $J_\mu$, associated with the global charge
appearing in eq.~\reef{charen}, \ie
 \beq
Q_\mt{A} =  \int_A d^{d-1}x\ J_{t}\,.
 \labell{ghost}
 \eeq
Again, symmetries and conservation of the current dictate the form of the
leading singularity in the correlator $\langle J_{\mu}\, \tsigma_n(\mue)
\rangle$. In this case, the singularity takes the form\footnote{Note that the
correlators here and in eq.~\reef{generalT} are implicitly evaluated in the
Euclidean path integral.}
 \be
\langle J_{i}\, \tsigma_n(\mue) \rangle = \frac{i\, k_n(\mue)}{2\pi}\,
\frac{\epsilon_{ij}\,n^j}{y^{d-1}}\,, \qquad \langle J_{a}\, \sigma_n\rangle =
0\,, \labell{generalJ}
 \ee
where $\epsilon_{ij}$ is the volume form in the two-dimensional space
transverse to $\Sigma$. This parity-odd tensor appears in the correlator
because of the magnetic flux carried by the generalized twist operator. We
refer to $k_n$ as the `magnetic response,' since this parameter characterizes
the response of the current to the magnetic flux.

Following \cite{renyi}, we can determine the value of $k_n$ using the conformal
mapping in the above discussion of spherical entangling surfaces. In this case,
one begins with the charge density that appears in the grand canonical ensemble
on the hyperbolic cylinder: $\langle J_{\taue}\rangle = -i\rho(n,\mue)$. Now
conformally mapping to the $n$-fold cover of $R^d$, this expectation value
becomes
 \be
\langle J_\mu\, \tsigma_n(\mue)\rangle_{flat} = \Omega^{d-2}\, \frac{\partial
X^{\alpha}}{\partial Y^{\mu}} \,\langle J_{\alpha}\rangle_{hyperbolic}\,.
 \labell{truck}
 \ee
The form of the transformation is fixed because the current has conformal
dimension $d$--1.\footnote{One can also verify that this transformation
\reef{truck} ensures that the charge operator \reef{ghost} defined on the
interior of the sphere, \ie $Q_\mt{A} = i \int_{\te=0,r<R} d^{d-1}x\, J_{\te}$,
is just the conformal transformation of the charge defined by integrating
$J_{\taue}$ over the entire hyperbolic plane $H^{d-1}$ --- as is implicit in
our discussions above. \labell{ice}} Now as indicated on the left-hand side of
eq.~\reef{truck}, this mapping yields the correlator of the current with the
spherical twist operator. Further, taking the limit where the current insertion
approaches the twist operator, one recovers the leading singularity in
eq.~\reef{generalJ}. Hence using eqs.~\reef{change} and \reef{factx}, the
magnetic response can be evaluated as
 \be
k_n(\mue)=2\pi n\,R^{d-1}\,\rho(n,\mue)\,.
 \labell{parmk}
 \ee
Here, the additional factor of $n$ appears because the correlators in
eq.~\reef{generalJ} are understood to involve the  the total current for the
entire $n$-fold replicated CFT whereas eq.~\reef{truck} corresponds to the
insertion of $J_\mu$ on a single sheet of the $n$-fold cover. Hence we must
multiply by an extra factor of $n$ to compare the two expressions.

An interesting universal property of $h_n$ was obtained for higher dimensional
twist operators in \cite{renyi,twistop} (see also \cite{Eric}):
 \be
 \partial_n h_n|_{n=1} = 2 \pi^{\frac{d}{2}+1}\,
 \frac{\Gamma \( {d}/{2}\)}{\Gamma(d+2)}\ C_T\,.
 \labell{interest1}
 \ee
Here $C_{T}$ is the central charge defined by the two-point function of the
stress tensor\footnote{Note that our normalization for $C_T$ here is a standard
one but it is not the same as in \cite{renyi}. Hence the numerical
factors in eq.~\reef{interest1} are slightly different than in that
reference.}
 \ba
\langle T_{\mu\nu}(x)T_{\rho\sigma}(0)\rangle={C_T\over
x^{2d}}\,\mathcal{I}_{\mu\nu,\rho\sigma}
 \labell{green}
 \eea
where
 \ba
\mathcal{I}_{\mu\nu,\rho\sigma}={1\over 2}
(I_{\mu\rho}I_{\nu\sigma}+I_{\mu\sigma}I_{\nu\rho})-{1\over
d}\delta_{\mu\nu}\delta_{\rho\sigma}\quad{\rm with}\ \ I_{\mu\nu}(x) =&&
\delta_{\mu\nu} - 2 \frac{x^\mu x^\nu}{|x|^2}\,.
 \labell{hot}
 \eea
In fact, similar universal properties is also found for higher derivatives of
$h_n$ in the vicinity of $n=1$.

The above universal behaviour does not immediately extend to the conformal
weight of the generalized twist operators $\tsigma(\mue)$. Instead, the natural
extension involves an expansion about both $n=1$ and $\mue=0$, as follows:
 \beq
 h_n(\mue)=\sum_{a,b} \frac{1}{a!\,b!}\ h_{ab}\ (n-1)^a\,\mue^b\qquad
 \labell{level}
 \eeq
where we defined the coefficients
 \beq
 h_{ab}\equiv(\partial_n)^a(\partial_{\mue})^b\, h_n(\mue)\big|_{n=1,\mue=0}\,.
 \labell{level2}
 \eeq
Note that the twist operator becomes trivial when $n=1$ and $\mue=0$ and hence
the first term in this expansion vanishes, \ie $h_{00}=0$. Further
$h_{10}=\partial_nh_n(\mue)\big|_{n=1,\mue=0}$ is precisely the term appearing
in eq.~\reef{interest1}. Now recall the expression \reef{weight} for the
weight, which we rewrite as
 \be
h_n(\mue) = \frac{2\pi n }{d-1}\,R^d\,\Big( \langle T_{\taue
\taue}\rangle\big|_{T_0/n,\mue}-\langle T_{\taue
\taue}\rangle\big|_{T_0,\mue=0}\Big)\,.\labell{weight2}
 \ee
in terms of the Euclidean stress tensor. Here both expectation values are in
the grand canonical ensemble on the hyperbolic space. That is, in terms of the
thermal density matrix given in eq.~\reef{eq:rho0x}, the density matrix
determining the second expectation value is $\rho_{\text{therm}}(\mue=0)$ while
that in the first is $[\rho_{\text{therm}}]^n$. Now we can produce the same
double expansion as in eq.~\reef{level} by re-expressing the latter with
 \be
\left[e^{-H/T_0 + \mue Q}\right]^n = e^{-H/T_0}\,\left[e^{-(n-1)H/T_0  + n\mue
Q}\right]
 \labell{level3}
 \ee
and expanding the last factor in terms of $n-1$ and $\mue$ \cite{twistop,Eric}.
Here, we note that the manipulation in eq.~\reef{level3} is valid since $Q$ is
a conserved charge and hence $[H,Q]=0$. Now it is straightforward to show that
the expansion coefficients $h_{ab}$ in eq.~\reef{level2} are given by
 \beq
h_{ab}= \frac{2\pi }{d-1}R^d (\partial_n)^a (\partial_{\mue})^b \, \left(n
\,\frac{Z(T_0,\mue=0)}{Z(T_0/n,\mue)}\langle T_{\taue\taue}e^{-(n-1) H/T_0 +
n\mue Q }\rangle - n \,\langle T_{\taue\taue}\rangle \right)\Big|_{n=1,\mue=0}
 \labell{hm}
 \eeq
Note that both of the expectation values above are evaluated in the thermal
ensemble with $T=T_0$ and $\mue=0$. Hence these coefficients can be determined
in terms of correlators of the stress tensor and the conserved current in the
thermal bath on the hyperbolic geometry at temperature $T_0$. However, by
applying the conformal transformation, these correlators may also be evaluated for
the CFT vacuum in flat space.

Let us focus here on the corrections to the conformal dimension of the twist
operator coming from a small chemical potential $\mue$  at $n=1$, \ie
 \beq h_{0b} =  \frac{2\pi \,R^d}{d-1}\,i^b
\langle\, T_{\taue\taue}\ \prod_{i=1}^b\int_{H^{d-1}}\!\!\!d^{d-1}\sigma_i \,
J_{\taue}(\sigma_i)\, \rangle_c
 \labell{hm20}
 \eeq
where the subscript $c$ denotes the connected correlator. Again, this
correlator is evaluated in the thermal ensemble on the hyperbolic space with
$T=T_0$ and $\mue=0$. However, as noted above, it is convenient to transform
back to flat space where the correlators will be evaluated in the CFT vacuum.
First, however, we observe that if we evaluate the correlator in
eq.~\reef{hm20} with a Euclidean path integral on $S^1\times H^{d-1}$, then the
stress tensor maybe inserted at any position in this background. Hence we
choose to place $T_{\taue\taue}$ at $(\taue,\sigma)=(\pi R,0)$ which the
conformal transformation then maps to $(\te,r)=(\infty,0)$ in the corresponding
flat space background. To be precise, with this choice, the conformal
transformation yields the following simple expression:
 \be
T_{\taue\taue} = \lim_{\te\to\infty}\left(\frac{\te^2}{2R^2}\right)^d
T_{\te\te} \,.
 \labell{transT}
 \ee
and so eq.~\reef{hm20} becomes
 \beq
h_{0b} =  \frac{2\pi \,R^d}{d-1}\,i^b \lim_{\te\to\infty}
\left(\frac{\te^2}{2R^2}\right)^d\langle\, T_{\te\te}\
\prod_{i=1}^b\int_{r<R}\!\!\!d^{d-1}x_i \, J_{\te}(x_i)\, \rangle_c
 \labell{hm2}
 \eeq
One immediate observation is that in the CFT vacuum in flat space, the
two-point correlator $\langle T J\rangle$ will vanish and hence we must have
$h_{01}=0$. That is, the linear correction in $\mue$ to the conformal weight of
the twist operator vanishes at $n=1$.

Hence the leading contribution should appear at order $\mue^2$ and is
determined by the three-point correlator $\langle T J J\rangle$. The latter
correlation function has a universal form dictated by conformal symmetry, up to
a few constants which are determined by the underlying CFT \cite{PetkouOsborn}.
Therefore $h_{02}$ can be determined entirely in terms of these few parameters.
More explicitly, the $\langle T J J\rangle$ correlator takes the following form
in a $d$-dimensional CFT \cite{PetkouOsborn}
 \be
\langle\,T_{\mu\nu}(x_1)\, J_\gamma(x_2)\, J_\delta(x_3)\,\rangle =
\frac{t_{\mu\nu\alpha\beta}({\bf X_{23}})\, I_{\gamma}{}^\alpha(x_{21})\,
I_{\delta}{}^\beta(x_{31})}{|x_{12}|^d |x_{13}|^d |x_{23}|^{d-2}} \,,
  \labell{3point}
 \ee
where
 \be
{\bf x_{12}}= {\bf x_1 - x_2}, \qquad {\bf X_{23}} = \frac{{\bf
x_{21}}}{|x_{21}|^2} - \frac{{\bf x_{31}}}{|x_{31}|^2}, \qquad \bf{\hat X} =
\frac{{\bf X}}{|X|},
 \ee
where $|{\bf x}|$ is the norm of the vector. Recall that $I_{\mu\nu}(x)$ was
defined in eq.~\reef{hot} and further we have
 \bea
t_{\mu\nu\rho\sigma}({\bf X}) &=& \ha\, h^1_{\mu\nu}({\bf\hat{X}})
\,\delta_{\rho\sigma} + \hb\,
h^1_{\mu\nu}({\bf\hat{X}})\,h^1_{\rho\sigma}({\bf\hat{X}}) + \hc\,
h^2_{\mu\nu\rho\sigma}({\bf\hat{X}}) + \he\,
h^3_{\mu\nu\rho\sigma}({\bf\hat{X}})\, ,
 \nonumber \\
h^1_{\mu\nu}({\bf\hat{ X}})&=& \hat{X}^\mu \hat{X}^\nu -
\frac{1}{d}\delta_{\mu\nu}\,,
 \label{aiai}\\
h^2_{\mu\nu\rho\sigma}({\bf\hat{ X}}) &=& \hat{X}^\mu \hat{X}^\rho\,
\delta_{\nu\sigma}+ \{\mu \leftrightarrow \nu,\rho \leftrightarrow \sigma \}
 \nonumber\\
&&\qquad - \frac{4}{d} \hat{X}^\mu \hat{X}^\nu \delta_{\rho\sigma} -
\frac{4}{d}\hat{X}^\rho\hat{X}^\sigma \delta_{\mu\nu} +
\frac{4}{d^2}\delta_{\mu\nu} \delta_{\rho\sigma}\,,
 \nonumber\\
h ^3_{\mu\nu\rho\sigma}({\bf\hat{ X}}) &=& \delta_{\mu\rho}\delta_{\nu\sigma} +
\delta_{\mu\sigma}\delta_{\nu\rho}
-\frac{2}{d}\delta_{\mu\nu}\delta_{\rho\sigma}\,.
 \nonumber
 \eea
The coefficients $\ha,\hb,\hc,\he$ are the parameters characterizing the
underlying CFT. However, only two of these constants are independent as they
satisfy the following constraints \cite{PetkouOsborn}:
 \be
d\,\ha - 2 \,\hb + 2(d-2)\, \hc=0\,, \qquad \hb- d(d-2)\, \he=0\,.
 \ee
Notice that in the special case $d=2$, both $\ha$ and $\hb$ vanish.

Now we only need to consider the correlator $\langle T_{\te \te}(x_1)
J_{\te}(x_2) J_{\te}(x_3)\rangle$ in the limit that $x^0_1\equiv\chi \to
\infty,\, x_1^i = 0$, while $x_2^0= x_3^0 = 0$ and  $|x_2|, |x_3| <R$. To
leading order in $\chi$, we find
 \ba
I_{00} &&= -1 +\cdots, \qquad I_{ij} = \delta_{ij} +\cdots, \qquad I_{i0} = \mathcal{O}(1/\chi)\\
 \eea
This immediately implies that for $\mu=\nu=\gamma=\delta=0$ in
eq.~\reef{3point}, we need only consider $t_{0000}$ to leading order and
further we have
 \be
t_{0000} \to \frac{1}{d^2} (-d\,\ha + \hb+ 4\,\hc + 2 d(d-1)\,\he)
=\frac{2}{d}\hc+\he \, .  \labell{t0000}
 \ee
Then for $h_{02}$ in eq.~\reef{hm2}, we are left with
 \ba
&&\langle T_{00}(x_1^0\to \infty)\, \int_{|x_2|<R}\!\!\! d^{d-1}x_2 J_0(x_2)
\,\int_{|x_3|<R}\!\!\! d^{d-1}x_3 J_0(x_3)\rangle    \nonumber \\
&&\qquad =\int_{|x_2|<R}\!\!\! d^{d-1}x_2\,\int_{|x_3|<R}\!\!\! d^{d-1}x_3\
\frac{\frac{2}{d}\hc+\he}{x_1^{2d} |x_{23}|^{d-2}}\,. \labell{Xint}
 \eea
The integral can be evaluated exactly. Using equation (\ref{hm2}), we finally
arrive at
 \be
%
%
h_{02} = -\frac{4\pi^{d-1}}{\Gamma(d+1)}\left(\frac{2}{d}\,\hc+\he\right)\,.
 \labell{BIGA}
 \ee

Now an analogous double expansion about $n=1$ and $\mue=0$ can also be applied
to the magnetic response:
 \beq
 k_n(\mue)=\sum_{a,b} \frac{1}{a!\,b!}\ k_{ab}\ (n-1)^a\,\mue^b\qquad
 \labell{levela}
 \eeq
where we defined the coefficients
 \beq
 k_{ab}\equiv(\partial_n)^a(\partial_{\mue})^b k_n(\mue)\big|_{n=1,\mue=0}\,.
 \labell{level2a}
 \eeq
Next we recall the expression \reef{parmk}, which we rewrite in terms of the
Euclidean current as
 \be
k_n(\mu) = 2\pi in\,R^{d-1}\,\langle
J_{\taue}\rangle\big|_{T_0/n,\mu}\,.\labell{parmk2}
 \ee

Again this expectation value is in the grand canonical ensemble on the
hyperbolic space. Now following the same manipulations of the corresponding
density matrix as in eq.~\reef{level3}, we arrive at the following expressions
for $k_{ab}$
 \beq
k_{ab}= 2\pi i \,R^{d-1}\, (\partial_n)^a (\partial_{\mue})^b \, \left(n
\,\frac{Z(T_0,\mue=0)}{Z(T_0/n,\mue)}\langle J_{\taue}e^{-(n-1) H/T_0  + n\mue
Q }\rangle  \right)\Big|_{n=1,\mue=0}
 \labell{hma}
 \eeq
where the remaining expectation value above is evaluated in the thermal
ensemble with $T=T_0$ and $\mue=0$. Hence these coefficients can again be
determined in terms of correlators of the stress tensor and the conserved
current in the thermal bath on the hyperbolic geometry at temperature $T_0$.
However, by conformally mapping to flat space, the correlators may
alternatively be evaluated in the CFT vacuum.

Let us evaluate a few coefficients for the low order contributions in the
expansion \reef{levela}. First, let us note that the coefficient $k_{10}$ will
determined in terms of the two-point correlator $\langle J T\rangle$ and so
upon mapping this correlator back to flat space, we will find a vanishing
result, \ie $k_{10}=0$. Considering the next two coefficients, eq.~\reef{hma}
yields
 \beqa
k_{01} &=& 2\pi i\,R^{d-1}\,\langle J_{\taue}\, Q  \rangle_c\,,
 \labell{km01}\\
k_{11} &=& -2\pi i\, R^{d-1} \left(\frac{1}{T_0} \langle J_{\taue}\, Q\,
H\rangle_c- 2\langle J_{\taue}\, Q \rangle_c\right)\,,
 \labell{km11}
 \eeqa
where subscript $c$ again denotes the connected correlators. These correlators
can be evaluated following the approach described above in evaluating $h_{02}$.
In particular, we conformally map these expressions back to flat space after
making a judicious choice for the position of the current insertion. The
resulting three-point function in eq.~\reef{km11} can be evaluated using the
$\langle T J J\rangle$ correlator given in eq.~\reef{3point}. Similarly, the
two-point function appearing in both expressions can be evaluated using the
current-current correlator
  \be
\langle J_\mu(x) J_\nu(0) \rangle =  \frac{C_{V}}{x^{2(d-1)}}\,I_{\mu\nu}(x)\,,
 \labell{twoJ}
 \ee
where $I_{\mu\nu}(x)$ was defined in eq.~\reef{hot}. Note that a Ward identity
relates the constant $C_V$ to the parameters appearing in the three-point
correlator \reef{3point} with \cite{PetkouOsborn}
 \beq
C_V =
\frac{2\pi^{d/2}}{\Gamma\left(\frac{d+2}2\right)}\left(\hc+\he\right)\,.
 \labell{extra}
 \eeq
Without discussing the calculations in more detail, let us present the
following results
 \beqa
\langle J_{\taue}\, Q  \rangle_c &=& - i
\frac{\pi^{(d-1)/2}}{2^{d-2}(d-1)\Gamma((d-1)/2)}\,\frac{C_{V}}{R^{d-1}}\,,
\nonumber\\
\langle J_{\taue}\, Q\, H \rangle_c &=&  - i \frac{2\pi^{d-2}}{d \Gamma(d-1)}\,
\frac{1}{R^{d}}\left(\frac{2}{d}\hc+\he\right)\,.
 \labell{poll}
 \eeqa
Substituting these results \reef{poll}, as well as eq.~\reef{extra}, into
eqs.~\reef{km01} and \reef{km11} then yields
 \beqa
k_{01} &=& \frac{8 \pi^d}{\Gamma(d+1)}(\hc+\he)\,,
 \labell{km01x}\\
k_{11} &=& \frac{8  \pi^d}{d\Gamma(d+1) } (2\hc - d (d-3) \he)\,.
 \labell{km11x}
 \eeqa
We might re-express the result for $k_{01}$ in a form similar to that appearing
in eq.~\reef{interest1} for the conformal weight, namely,
 \be
 \partial_{\mue} k_n(\mue)|_{n=1,\mue=0} =   4\pi^{d/2}
 \frac{\Gamma\left(\frac{d+2}2\right)}{\Gamma(d+1)}\,C_V\,,
 \labell{interest2}
 \ee
where $C_V$ is the central charge appearing in the current-current correlator
\reef{twoJ}.

\subsection{Generalized twist operators in $d$=2}
\labell{Gen twist op d2}

In this subsection, we compute  charged \ren entropies using twist operators in
a simple two-dimensional CFT.\footnote{The analysis in this section was first
done by T. Takayanagi \cite{tadashi-twist}. We thank him for sharing these
results with us.} In particular, we consider a free massless Dirac fermion
$\psi$ on an infinite line and we are interested in the \ren entropy of a
subsystem $x\in[u,v]$. In accord with the review at the beginning of this
section, the \ren entropy can determined by evaluating the partition function
of $\psi$ on an $n$-sheeted cover of $R^2$, which is equivalent to the
correlation function of twist operators inserted at the entangling surface, \ie
the two points $x=u,\,v$ \cite{cardy0}. Let us first review the computation of
the free fermion without the Wilson loop, as in \cite{ant}. On a $n$-fold
cover, there is a branch cut connecting $x=u$ and $v$ and each time we cross
the branch cut, we change from one sheet to the next. Let us label the fermion
on $k$-th sheet as $\psi_k$, where $k$ runs from $1$ to $n$. Then the fields on
the different sheets are identified as follows:
 \ba
\psi_{k}(e^{2\pi i}(w-u))=\psi_{k+1}(w-u)\,,\qquad\psi_{k}(e^{2\pi
i}(w-v))=\psi_{k-1}(w-v)\,,
 \labell{replica-ident-noWilson}
 \eea
where we used the complexified coordinate $w=x+i\te$. These boundary conditions
can be `diagonalized' by defining $n$ new fields
 \ba
\tilde{\psi}_m={1\over n}\sum_{k=1}^{n}e^{2\pi i k m/n}\,\psi_{k}
 \labell{diagP}
 \eea
for which the boundary conditions \reef{replica-ident-noWilson} become
 \ba
\tilde{\psi}_m(e^{2\pi i}(w-u))=e^{2\pi i m/n}\tilde{\psi}_m(w-u)\,,~
 \tilde{\psi}_m(e^{2\pi i}(w-v))=e^{-2\pi
i m/n}\tilde{\psi}_m(w-v)
 \labell{replica-diag-noWilson}
 \eea
where $m=-(n-1)/2,-(n-1)/2+1,\cdots, (n-1)/2$. The phase shifts in
eq.~(\ref{replica-diag-noWilson}) are generated by standard twist operators
$\sigma_{m/n}$, each of which act only on the corresponding $\tilde{\psi}_m$
and which have conformal dimension $\Delta_m={1\over 2}(m/n)^2.$ The full twist
operator $\sigma_n$ appearing in evaluating the \ren entropy can then be
written as $\sigma_n=\prod \sigma_{m/n}$ and hence the desired correlator of
the twist operators $\sigma_{n}$ and $\sigma_{-n}$ yields:
 \ba
Z_n=\langle\sigma_{n}(u)\,\sigma_{-n}(v) \rangle= \prod_{m=-{n-1\over
2}}^{{n-1\over 2}}\langle\sigma_{m/n}(u)\,\sigma_{-m/n}(v) \rangle \sim
|u-v|^{-4\Delta_{n}} \,,
 \labell{chewb}
 \eea
where total conformal dimension $\Delta_{n}$ appearing above is given by
 \ba
\Delta_{n}=\sum_{m=-{n-1\over 2}}^{{n-1\over 2}}{1\over 2}\left({m\over
n}\right)^2={1\over 24}\left(n-{1\over n}\right) \,.
 \labell{totalD}
 \eea
Then applying eq.~\reef{ren2} to evaluate the \ren entropy, we recover the
well-known result
 \ba
S_{n}={1\over 6}\left(1+{1\over n}\right) \log |u-v|\,.
 \labell{hurrah}
 \eea

We now generalize the above discussion to evaluate the charged \ren entropy. In
particular, the charge, which we consider here, will be that associated with
global phase rotations of the fermion, $\psi\to e^{i\theta}\psi$. If we
consider an imaginary chemical potential $\imu$, the effect of the Wilson loop
is easily represented by extending the original boundary conditions
\reef{replica-ident-noWilson} to include a additional phase:
 \ba
\psi_{k}(e^{2\pi i}(w-u))=e^{i\imu}\psi_{k+1}(w-u)\,,\qquad\psi_{k}(e^{2\pi
i}(w-v))=e^{-i\imu}\psi_{k-1}(w-v)
 \labell{replica-ident}
 \eea
Since this additional phase is added uniformly, the `diagonal' fields
\reef{diagP} now satisfy
 \beqa
\tilde{\psi}_m(e^{2\pi i}(w-u))&=&e^{2\pi i m/n+i\imu}\tilde{\psi}_m(w-u)
\,,\nonumber\\
 \tilde{\psi}_m(e^{2\pi i}(w-v))&=&e^{-2\pi
i m/n-i\imu}\tilde{\psi}_m(w-v)\,.
 \labell{diagBC}
 \eea
These phase shifts are accomplished by introducing twist operators,
$\sigma_{\alpha(m,\imu)}$ and $\sigma_{-\alpha(m,\imu)}$, where
 \beq
\alpha(m,\imu)=\frac{m}{n}+\frac{\imu}{2\pi}+\ell_m\,,
 \labell{alphaX}
 \eeq
where $m$ runs from $-{n-1\over 2}$ to ${n-1\over 2}$ as, before. The conformal
dimension of these twist operators is now
 \ba
\Delta_{\alpha(m,\imu)}={1\over 2}\alpha(m,\imu)^2={1\over 2}\left({m\over
n}+\frac{\imu}{2\pi}+\ell_m\right)^2\,.
 \labell{newD}
 \eea
The constant $\ell_m$ appearing above is an integer which is chosen to minimize
the conformal dimension of the corresponding twist operator. This freedom
arises because of the ambiguity in defining the phase factors in
eq.~\reef{diagBC} modulo $2\pi$. For example, shifting $\ell_m$ from 0 to 1
changes the corresponding phase factor by $2\pi$ and so leaves the
corresponding boundary condition in eq.~(\ref{diagBC}) unchanged. The conformal
dimension \reef{newD} is always minimized by choosing $\ell_m$ so that the
phase factor generated by the twist operator lies between $-\pi$ and $\pi$, \ie
such that $-\frac12\le \alpha(m,\imu) \le \frac12$.

\begin{figure}[h!]
\begin{center}
\includegraphics[width=\textwidth]{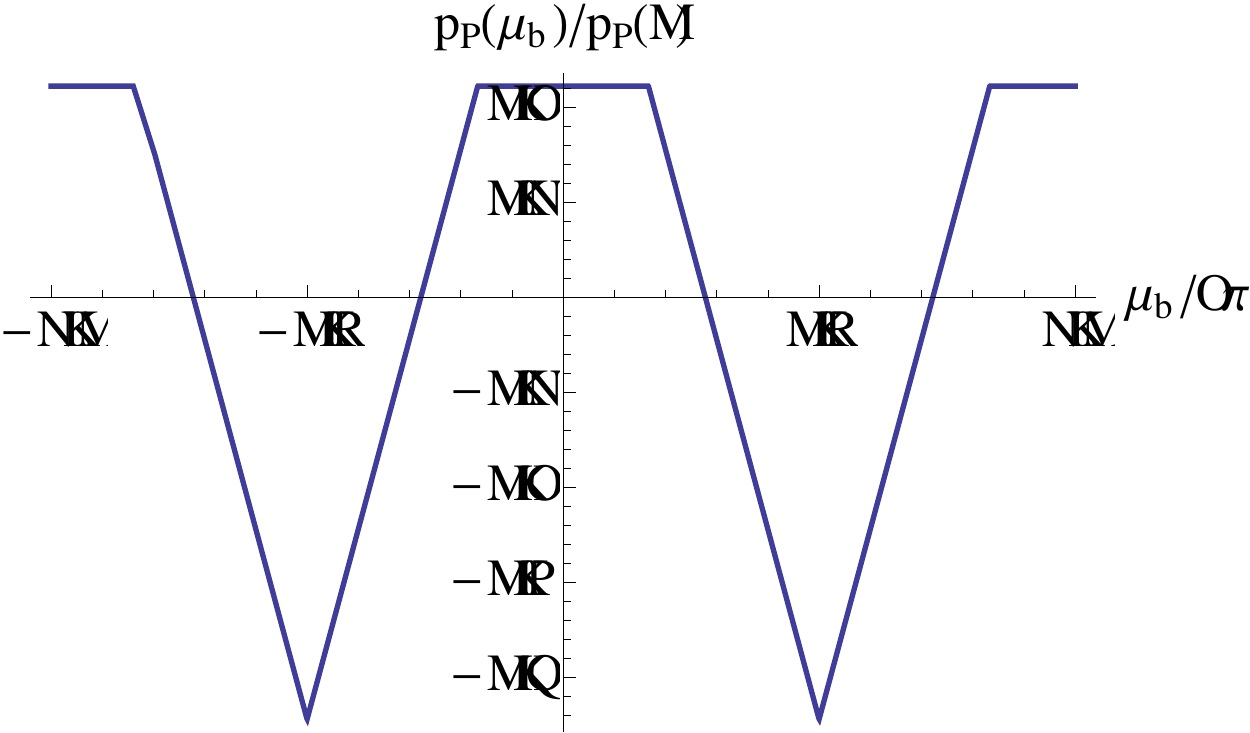}
\caption{Charged \ren entropy with $n=3$ for a two-dimensional free fermion as
a function of $\imu$.} \label{twistfreeferm}
\end{center}
\end{figure}

When $\imu$ is small enough that all of the phase factors lie between $-\pi$
and $\pi$, \ie
 \ba
\Big|{m\over n}+\frac{\imu}{2\pi}\Big|\le {1\over 2}\quad{\rm for}\ \ m\in
\left[-{n-1\over 2}, {n-1\over 2}\right]\,,
 \labell{boundm}
 \eea
we will have $\ell_m=0$ for all $m$. If we assume the latter holds, the
conformal dimension of the generalized twist operator
$\tilde\sigma_n=\prod\tilde\sigma_{\alpha(m,\imu)}$ becomes
 \ba
\Delta_{n}={1\over 24}\left(n-{1\over n}\right)+{n \over 2 }\,\left({\imu \over
2\pi }\right)^2 \,,
 \labell{outside}
 \eea
and the charged \ren entropy is given by
 \ba
\tS_{n}(\imu)={1\over 6}\left(1+{1\over n}\right) \log |u-v|\,.
 \label{replica-renyi}
 \eea
That is, for small $\imu$, the charged \ren entropy is exactly the same as the
result in eq.~\reef{hurrah}, \ie the \ren entropy without the Wilson line. As
$\imu$ increases, we can no longer choose all of the $\ell_m$ to be zero. The
first transition occurs for $m=\frac{n-1}2$ when
 \ba
{n-1\over 2n}+\frac{\imu}{2\pi}= {1\over 2} \quad\longleftrightarrow \quad \imu
={\pi\over n}
 \labell{amber}
 \eea
beyond which the naive phase factor would be larger than $\pi$. Setting
$\ell_{\frac{n-1}{2}}=-1$ and using the appropriate conformal dimension, we
find that the charged \ren entropy within the range ${\pi\over n}\le \imu <
{3\pi\over n}$ becomes
 \ba
\tS_{n}(\imu)=\left[{1\over 6}\left(1+{1\over n}\right) -{4\over
n-1}\left(\frac{\imu}{2\pi}-{1\over 2n}\right)\right] \log |u-v|\,.
 \label{replica-renyi-line}
 \eea
Further phase transitions occur whenever $\imu=\frac{\pi}{n}(2k+1)$. For
example, for ${3\pi\over n}\le \imu < {5\pi\over n}$, the charged \ren entropy
becomes
 \ba
 \tS_{n}(\imu)=\left[{1\over 6}\left(1+{1\over n}\right) +{4\over n-1}\left(\frac{\imu}{2\pi}
 -{5\over 2n}\right)\right] \log |u-v|\,.
 \label{replica-renyi-line2}
 \eea
Of course, it is straightforward to extend these results to all values of
$\imu$. As can be anticipated from eq.~\reef{replica-ident}, the charged \ren
entropy exhibits a periodicity
 \bea
 \tS_{n}(\imu)=\tS_{n}(\imu+2\pi)\,.
 \labell{waveX}
 \eea
Hence within a single period, there will be $n$ separate branches running from
$\imu=\frac{\pi}{n}(2k-1)$ to $\frac{\pi}{n}(2k+1)$ for integer $k$. The result
for $n=3$ is shown in figure \ref{twistfreeferm}. Of course, these results show
that the charged \ren entropy is a non-analytic function of $\imu$ and $n$. In
particular, we might note that the apparent singularities at $n=1$ in
eqs.~\reef{replica-renyi-line} and \reef{replica-renyi-line2} are not physical.
As a final comment, we remark that the results derived here using twist
operators agree with those coming from the heat kernel computations in Appendix
\ref{free}.

%


\section{Holographic computations}\labell{Einstein}

In this section we calculate holographic \ren entropies for boundary theories
dual to Einstein gravity coupled to a Maxwell gauge field in the bulk. The
relevant bulk solutions are charged topological black holes with hyperbolic
horizons. These solutions represent the grand canonical ensemble of the
boundary CFT on the hyperbolic cylinder. We only present the salient steps in
the following calculations and refer the reader to \cite{renyi} for a detailed
description of how the holographic \ren entropies are calculated. We consider
boundary theories in dimension $d\ge3$ here and provide holographic
calculations for $d=2$ in appendix \ref{threed}.

\subsection{Charged black hole solution}

In $d+1$ bulk dimensions, we write the Einstein-Maxwell action with negative
cosmological constant as\footnote{The scale $\lstar$ appearing in the prefactor
of the Maxwell term should be fixed by the details of the boundary theory. With
this notation, the ($d$+1)-dimensional gauge coupling becomes
$g_5^{\,2}=2\lp^{d-1}/\lstar^2$.}
\begin{equation}
I_{E-M} = \frac{1}{2\lp^{d-1}} \int d^{d+1}x\,\sqrt{-g}\left(\frac{d(d-1)}{L^2} + \mathcal{R}
-\frac{\lstar^2}{4} F_{\mu\nu}F^{\mu\nu}\right)\,.
\labell{action}
\end{equation}
For $d\ge3$, the metric for the charged topological black hole takes the form
\begin{equation}
ds^2 = -f(r) \frac{L^2}{R^2}\, d\tau^2 +
\frac{dr^2}{f(r)} + r^2\, d\Sigma_{d-1}^2\,,
\labell{bhmetric}
\end{equation}
with
\begin{equation}
f(r) = \frac{r^2}{L^2}-1-\frac{m}{r^{d-2}}+\frac{q^2}{r^{2d-4}}
\labell{funct}
\end{equation}
where $d\Sigma_{d-1}^2 = du^2+\sinh^2\!u\, d\Omega^2_{d-2}$ is the metric on
$H^{d-1}$ with unit curvature. Note that the time coordinate is normalized here
\cite{casini9} so that the boundary metric naturally becomes $ds^2_{CFT} =
-d\tau^2+R^2d\Sigma_{d-1}^2$, \ie the Minkowski continuation of
eq.~\reef{frog}. The corresponding bulk gauge field is
\begin{equation}
A = \left( \sqrt{\frac{2(d-1)}{(d-2)}}\frac{L\,q}{R\lstar\, r^{d-2}}-\frac{\mue}{2\pi R}
\right)d\tau\,,
\labell{gauge}
\end{equation}
The chemical potential $\mue$ is fixed by requiring that the gauge field vanish
at the horizon  $r=r_H$, \ie
\begin{equation}
\mue = 2\pi\sqrt{\frac{2(d-1)}{(d-2)}}\frac{L\,q}{\lstar r_H^{d-2}} \,.
\end{equation}
The mass parameter $m$ is related to the horizon radius $r_H$ by
\begin{equation}
m = \frac{r_H^{d-2}}{L^2}(r_H^2-L^2)+\frac{q^2}{r_H^{d-2}}  \,.\labell{mass parameter}
\end{equation}
Hence, we may rewrite the function $f(r)$ (\ref{funct}) in terms of the horizon
radius $r_H$ and the charge $q$, giving
\begin{equation}
f(r) = \frac{r^2}{L^2} - 1 + \frac{q^2  }{r^{2d-4}} - \left(\frac{r_H}{r}\right)^{d-2} \left(\frac{r_H^2}{L^2}-1 +
\frac{q^2}{r_H^{2d-4}}\right) \,.
\end{equation}
The temperature of this black hole is given by
\begin{equation}
T = \frac{T_0}2\, L f'(r_H)=\frac{T_0}2\, \left[
d\,\frac{r_H}{L}-(d-2)\frac{L}{r_H}\left(1+\frac{d-2}{2(d-1)}\left(
\frac{\mu\,\lstar}{2\pi L}\right)^2\right)\right]
\labell{gambit}
\end{equation}
where $T_0$ is the temperature given in eq.~\reef{bus} and the `prime' denotes
differentiation with respect to $r$. The thermal entropy is given by the
Bekenstein-Hawking formula
\begin{equation}
S = \frac{2 \pi}{\lp^{d-1}}  V_\Sigma\,r_H^{d-1}\,, \labell{thermal}
\end{equation}
where $V_\Sigma$ denotes the regulated (dimensionless) volume of the hyperbolic
plane $H^{d-1}$, as described in \cite{renyi}. Recall that this volume is a
function of $R/\delta$, the ratio of the radius of the entangling sphere to the
short-distance cut-off in the boundary theory. Further the leading contribution
takes the form
 \be
V_\Sigma \simeq
\frac{\Omega_{d-2}}{d-2}\,\frac{R^{d-2}}{\delta^{d-2}}+\cdots\,,
 \labell{led99}
 \ee
where $\Omega_{d-2}=2\pi^{(d-1)/2}/\Gamma((d-1)/2)$ is the area of a unit
($d$--2)-sphere. Hence the corresponding \ren entropies in the following begin
with an area law contribution.

As a final comment, we note that we have presented the Minkowski-signature
solution here with a real chemical potential $\mue$. This gives the holographic
representation of the grand canonical ensemble on the hyperbolic cylinder
$R\times H^{d-1}$. One can easily transform to Euclidean signature by replacing
$\tau=-i\taue$ to produce the dual of the thermal ensemble for the boundary CFT
on $S^1\times H^{d-1}$. In this replacement, the form of the metric function
$f(r)$ is unchanged and as usual, the Euclidean time is made periodic with
$\Delta\taue=1/T$ to ensure that the bulk geometry is smooth at $r=r_H$. 
The gauge field becomes imaginary for this Euclidean bulk solution. Of
course, the latter is in keeping with our discussion of the Euclidean path
integral in section \ref{two}, where an imaginary background gauge field was
introduced to describe the grand canonical ensemble. Here, this background
field in the boundary theory is simply given by the non-normalizable of the
bulk gauge field, \ie $B_\mu = -\lim_{r\to\infty}A_\mu$.

\subsection{Charged \ren entropies}

Applying eq.~(\ref{Rnyi-therm}) with a real chemical potential, we see that the
charged \ren entropy for a spherical entangling surface can be expressed as
\begin{equation}
S_{n}(\mue) = \frac{n}{n-1} \frac{1}{T_0}\int^{x_1}_{x_n} S(x,\mue)\,\partial_xT(x,\mue)\,dx
 \,, \labell{ent9}
\end{equation}
where $x = r_H/L$ and $S(x,\mue)$ is the horizon entropy \reef{thermal}.
Evaluating eq.~\reef{gambit} in terms of $x$ gives
\begin{equation}
T(x,\mue) = \frac{T_0}{2 x} \left(dx^2-(d-2) -
 \frac{(d-2)^2}{2(d-1)}\left(\frac{\mue \lstar}{2\pi L}\right)^2  \right)\,.
\end{equation}
Then $x_n$ is the largest solution of $T(x_n,\mue)=T_0/n$ and is given by
\begin{equation}
x_n = \frac{1}{dn} +
\sqrt{\frac{1}{d^2n^2}+ \frac{d-2} d+\frac{(d-2)^2}{2d(d-1)}
\left(\frac{\mue \lstar}{2\pi L}\right)^2}\,.
 \labell{joke}
\end{equation}
Combining these expressions then yields
 \beqa
S_{n}(\mue) &=&   \pi V_\Sigma \left(\frac{L}{\lp}\right)^{d-1}\frac{n}{n-1}
\left[ \left(1+\frac{d-2}{2(d-1)}\left(\frac{\mue \lstar}{2\pi L}\right)^2\right )
(x_1^{d-2}-x_n^{d-2}) + x_1^d-x_n^d\right] \,.\nonumber\\ && \labell{renyi entropy}
 \eeqa
Note that when $\lstar=0$, eqs.~\reef{joke} and \reef{renyi entropy} reduce to
the results found in \cite{renyi}.

Expressions for the charged \ren entropy with specific choices for $n$ are:
 \beqa
\lim_{n\to0} S_n &=&\ \  \pi V_\Sigma \left(\frac{L}{\lp}\right)^{d-1}
 \left(\frac{2}{d}\right)^d \frac{1}{ n^{d-1}} \labell{bell}\\
S_\mt{EE}=\lim_{n\to1}S_n &=&\ \ \pi V_\Sigma \left(\frac{L}{\lp}\right)^{d-1}
\frac{(d-2)x_1^{d-2}}{d x_1-1}\left(
1+\frac{d-2}{2(d-1)}\left(\frac{\mue \lstar}{2\pi L}\right)^2
+\frac{d\,x_1^2}{d-2}\right) \nonumber\\
S_2 &=& 2 \pi V_\Sigma \left(\frac{L}{\lp}\right)^{d-1}
 \left( \left(1+\frac{d-2}{2(d-1)}\left(\frac{\mue \lstar}{2\pi L}\right)^2\right)
(x_1^{d-2}-x_2^{d-2}) +  x_1^d-x_2^d\right) \nonumber\\
\lim_{n\to\infty} S_n &=&\ \ \pi V_\Sigma \left(\frac{L}{\lp}\right)^{d-1}
\left( \left(1+\frac{d-2}{2(d-1)}\left(\frac{\mue \lstar}{2\pi L}\right)^2\right)
(x_1^{d-2}-x_\infty^{d-2}) +  x_1^d-x_\infty^d\right) \nonumber
 \eeqa
where
\begin{equation}
x_\infty^{\,2} =\frac{d-2}{d}
\left(1+\frac{d-2}{2(d-1)}\left(\frac{\mue \lstar}{2\pi L}\right)^2\right)\,.
\end{equation}
Another interesting limit to consider is holding $n$ fixed while $\mue
\rightarrow \infty$, which yields
\begin{equation}
\lim_{\mue \rightarrow \infty} S_n(\mue) = 2 \pi V_\Sigma \left(\frac{(d-2)^2}{2d(d-1)}\right)^{\frac{d-1}{2}}
 \left(\frac{\lstar}{\lp}\right)^{d-1}\left(\frac{\mu}{2 \pi}\right)^{d-1}\,.
 \labell{curiousX}
\end{equation}
Hence we have the curious result that, to leading order in $\mue$, the \ren
entropies are independent of $n$ in this limit.

\begin{figure}[h!]
\centering
\begin{subfigure}[b]{0.49\textwidth}
\caption{}
\centering
\includegraphics[width=\textwidth]{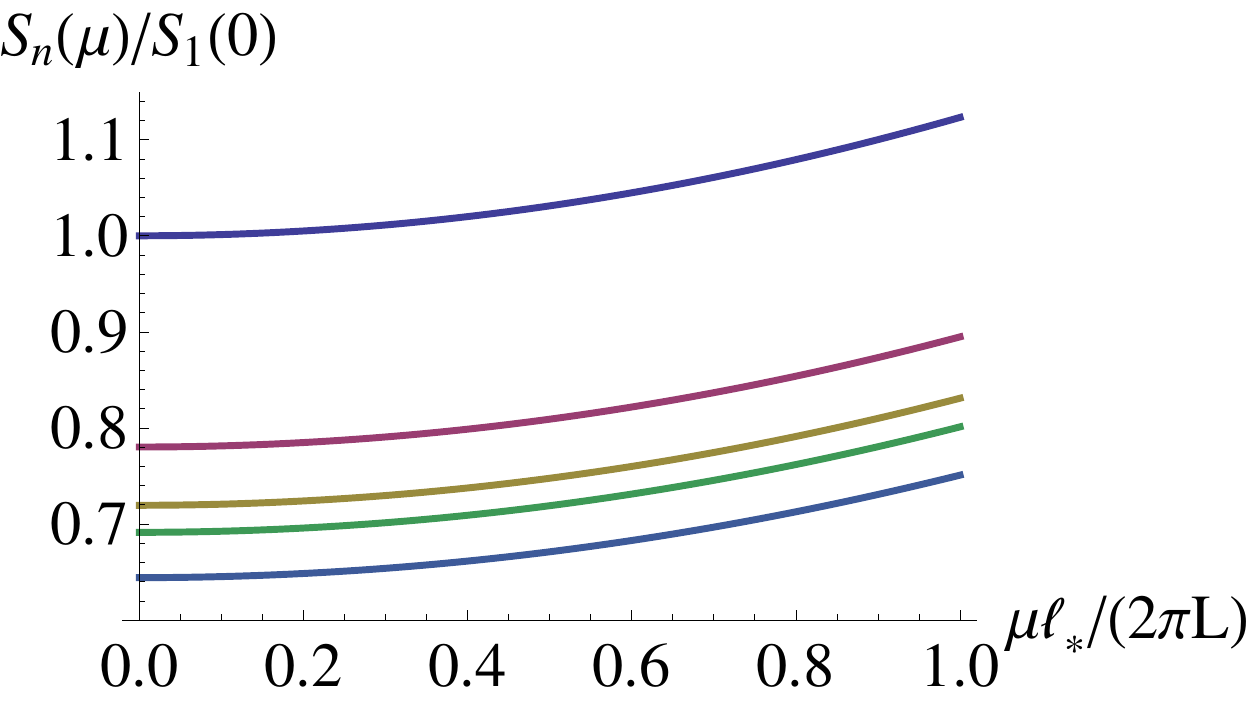}
\end{subfigure}
\begin{subfigure}[b]{0.49\textwidth}
\caption{}
\centering
\includegraphics[width=\textwidth]{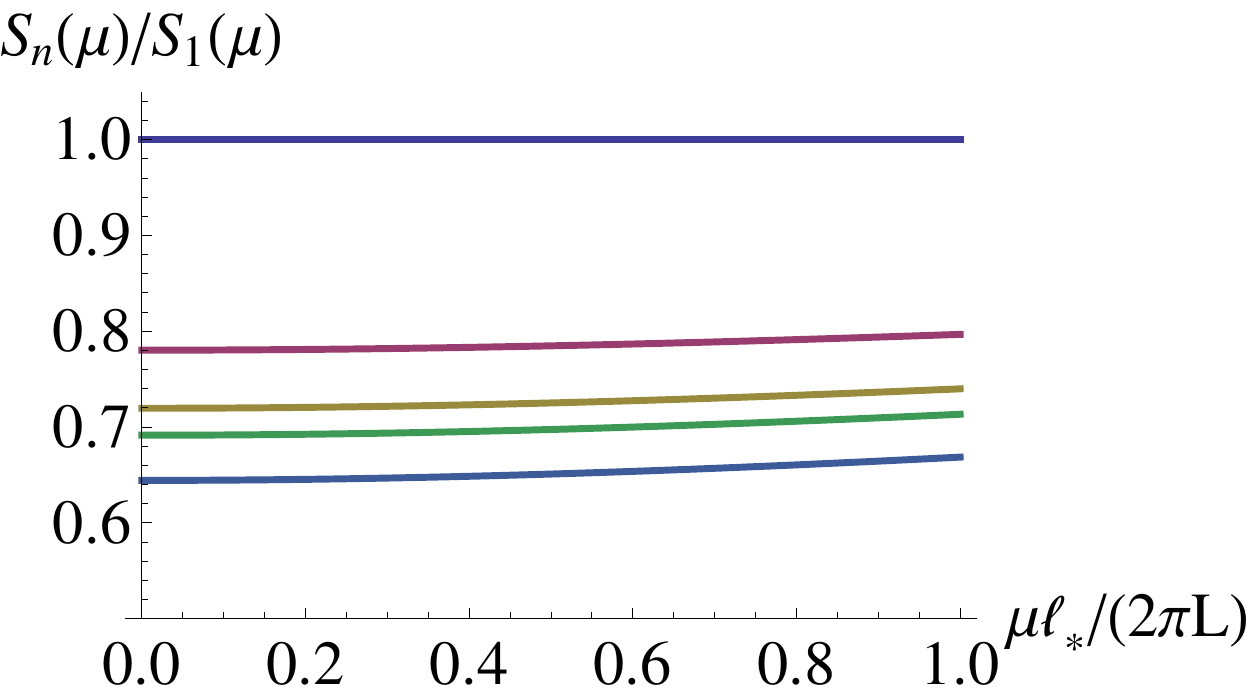}
\end{subfigure}
\caption{
The $d=3$ charged \ren entropy (normalized by (a) $S_1(0)$ and (b) $S_1(\mue)$) as a
function of $\mue$. The curves correspond to (from top to bottom) $n$=1,2,3,4,10}
\label{three}
\end{figure}
\begin{figure}[h!]
\centering
\begin{subfigure}[b]{0.49\textwidth}
\caption{}
\centering
\includegraphics[width=\textwidth]{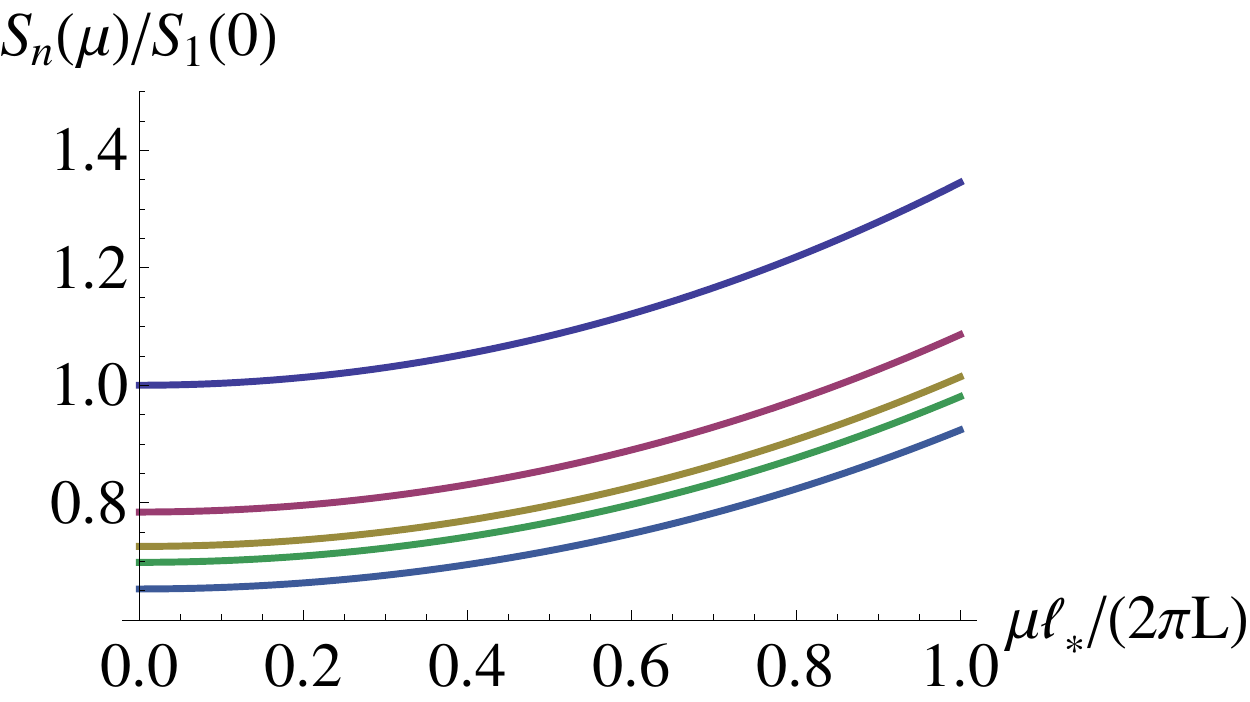}
\end{subfigure}
\begin{subfigure}[b]{0.49\textwidth}
\caption{}
\centering
\includegraphics[width=\textwidth]{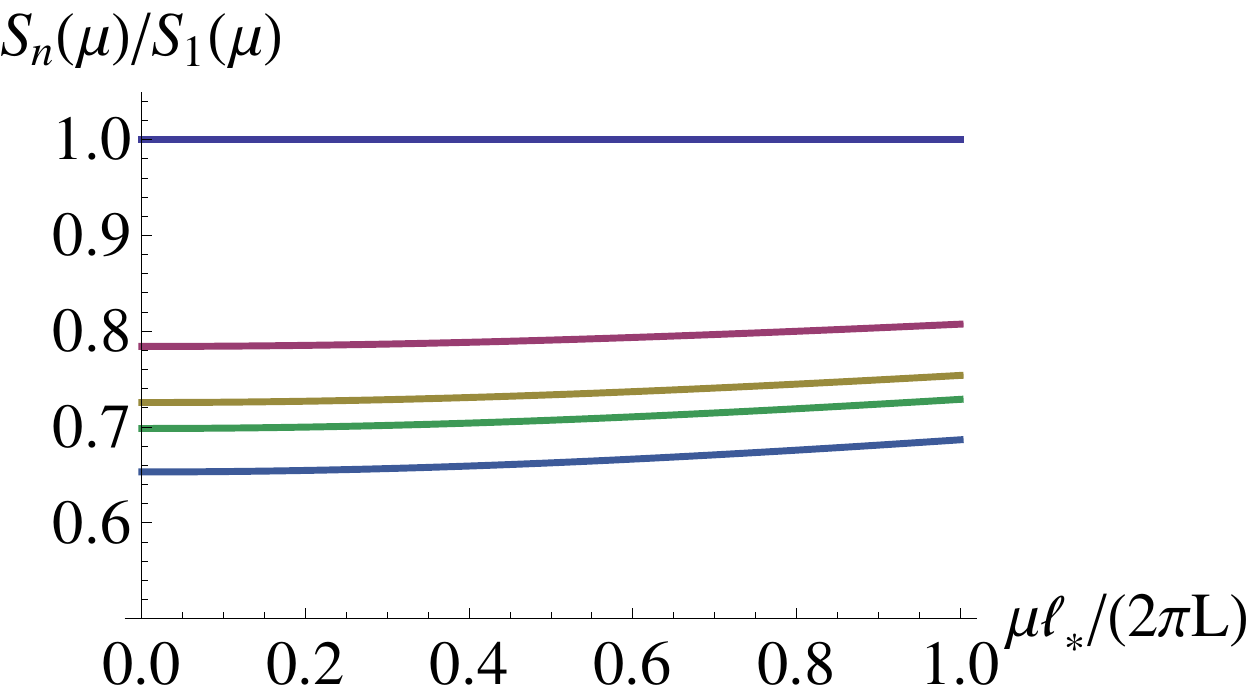}
\end{subfigure}
\caption{The $d=4$ charged \ren entropy (normalized by (a) $S_1(0)$ and (b) $S_1(\mue)$) as a
function of $\mue$. The curves correspond to (from top to bottom) $n$=1,2,3,4,10}
\label{four}
\end{figure}
\begin{figure}[h!]
\centering
\begin{subfigure}[b]{0.49\textwidth}
\caption{}
\centering
\includegraphics[width=\textwidth]{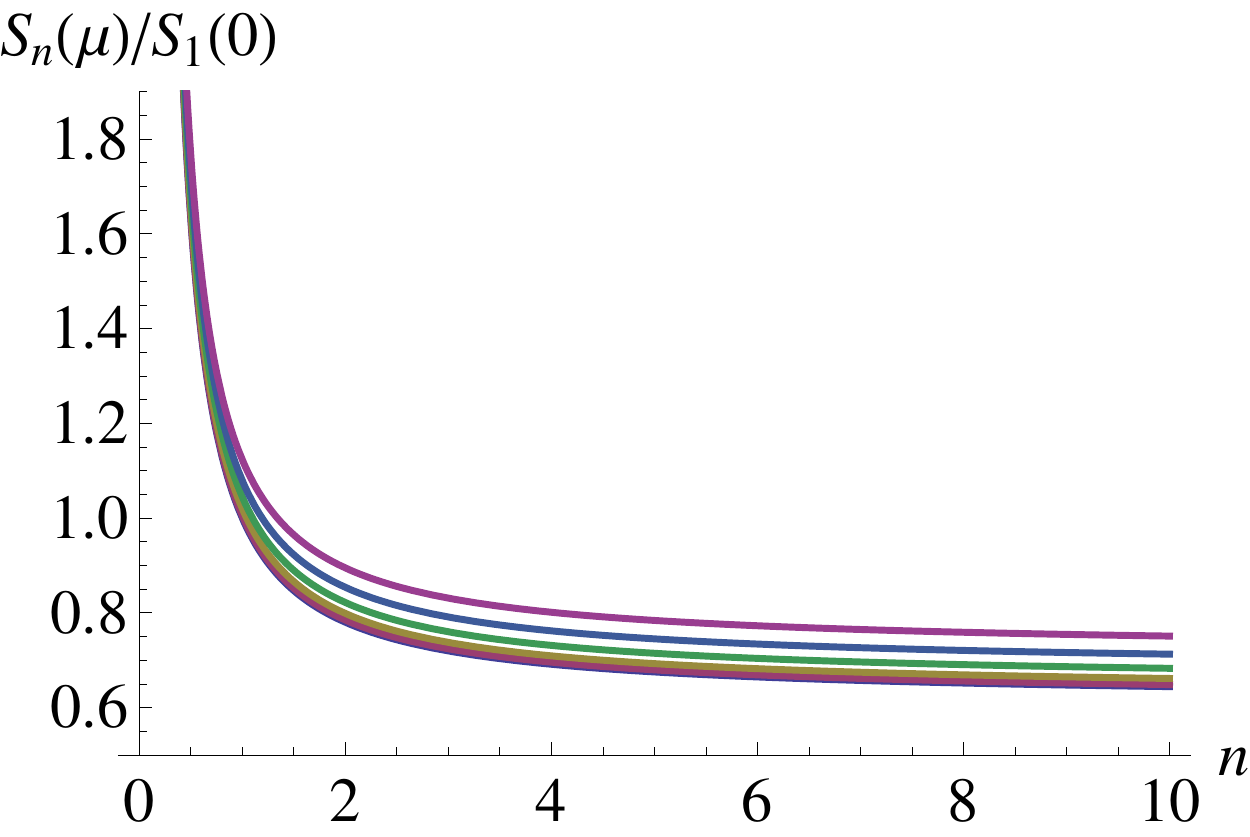}
\end{subfigure}
\begin{subfigure}[b]{0.49\textwidth}
\caption{}
\centering
\includegraphics[width=\textwidth]{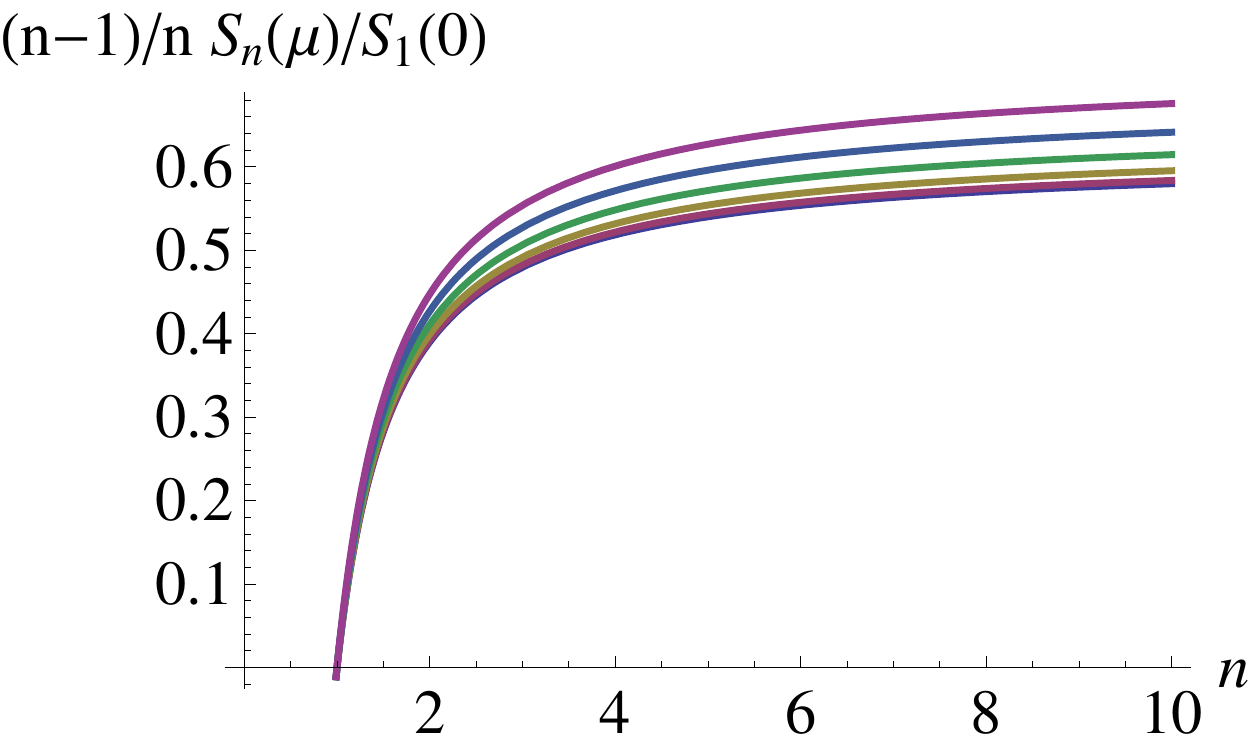}
\end{subfigure}
\caption{The charged \ren entropy (normalized by $S_1(0)$) in $d=3$ shown
function of $n$ in panel (a). In panel (b), we show $\frac{n-1}{n}S_n(\mue)$
as a function of $n$. Note
that the slope of the curves is negative in panel (a) and positive in panel
(b). In both cases, the curves correspond to (from top to bottom)
$\frac{\mue \lstar}{2 \pi L}=1.0,0.8,0.6,0.4,0.2$
and $0.0$.} \label{fourn}
\end{figure}

The results for the charge \ren entropy \reef{renyi entropy} are illustrated in
figures \ref{three} and \ref{four}, which plot $S_{n}(\mue)$ as a function of
$\mue$ for various values of $n$ in $d=3$ and 4. In these figures, it is
evident that for fixed $\mue$, the \ren entropy decreases as $n$ increases.
This behaviour is also shown in figure \ref{fourn}a where the charged \ren
entropy in $d=3$ is shown as a function of $n$. The figure shows very clearly
in this example that $\partial_n S_n(\mue)<0$. As discussed in \cite{renyi}
(see also \cite{renyi1}), standard \ren entropies must satisfy various
inequalities:
 \beqa
\frac{\partial S_n}{\partial n} \le 0\,,&&\qquad\qquad \frac{\partial}{\partial
n}\left(\frac{n-1}{n} S_n \right) \ge  0\,,
\labell{inequalities}\\
\frac{\partial}{\partial n}((n-1)S_n)\ge0\,,&&\qquad\qquad
\frac{\partial^2}{\partial n^2}((n-1)S_n)\le 0\,. \nonumber
 \eeqa
By examining plots of the numerical results, \eg see $\frac{n-1}{n}S_n(\mue)$
in figure \ref{fourn}b, we find that these inequalities still appear to hold in
the charged case. The analysis in \cite{renyi} found that this result
essentially follows from the connection between the \ren entropies for a
spherical entangling surface and the thermal entropy on the hyperbolic cylinder
$R\times H^{d-1}$. In particular, it follows that these inequalities
\reef{inequalities} will be satisfied for any CFT, as long as the corresponding
thermal ensemble is stable. Thus we expect that the inequalities
(\ref{inequalities}) will continue to hold for charged \ren entropies. Of
course, the arguments in \cite{renyi} will not apply where the \ren entropies
exhibit phase transitions \cite{renyiphases}, as discussed in section
\ref{stability}.

\begin{figure}[h!]
\centering
\begin{subfigure}[b]{0.49\textwidth}
\caption{}
\centering
\includegraphics[width=\textwidth]{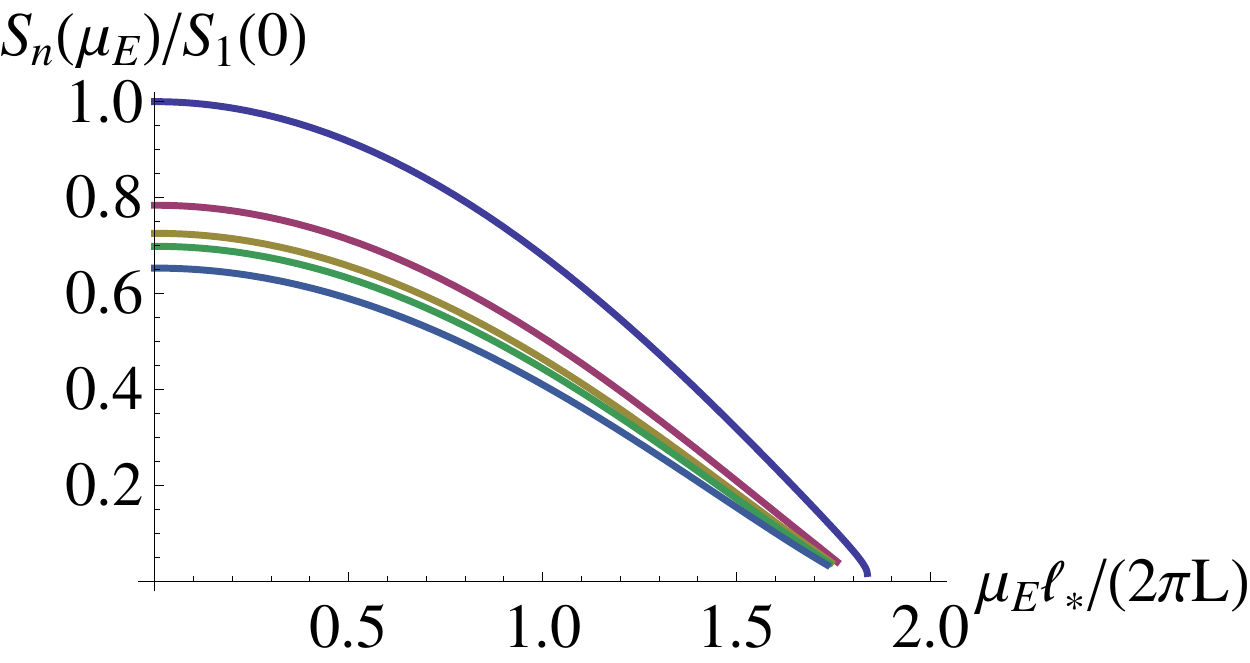}
\end{subfigure}
\begin{subfigure}[b]{0.49\textwidth}
\caption{}
\centering
\includegraphics[width=\textwidth]{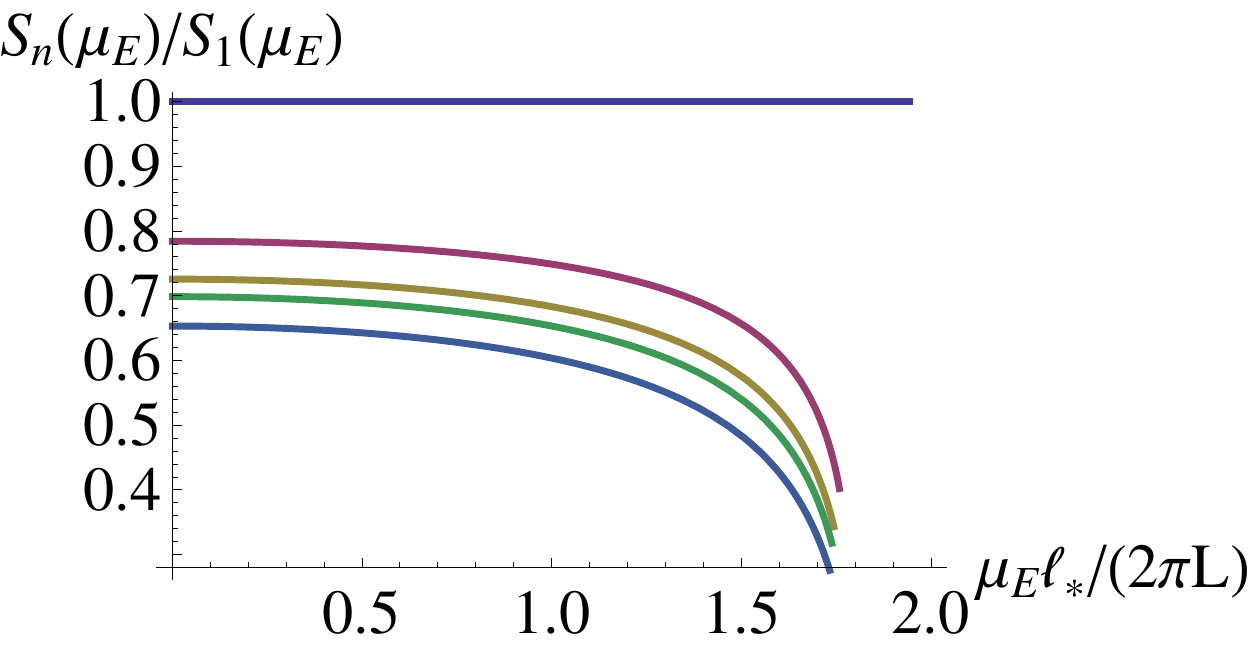}
\end{subfigure}
\caption{
The $d=4$ charged \ren entropy (normalized by (a) $\tS_1(0)$ and (b)
$\tS_1(\imu)$) as a function of the imaginary chemical potential $\imu$. The
curves correspond to (from top to bottom) $n$=1,2,3,4,10} \label{imthree}
\end{figure}

As we mentioned above, both real and imaginary chemical potentials are of
interest. Our holographic results are easily analytically continued to
imaginary chemical potential by simply replacing $\mue=i\imu$ and $q=i\iq$.
Note that with this replacement, the root $x_n$ in eq.~\reef{joke} fails to
exist if $\imu$ becomes too large. The region of validity of analytically
continued solutions is given by
\begin{equation}
 \imu^2 \le \frac{8\pi^2(d-1)}{d-2} \left(\frac{L}{\lstar}\right)^2\left(1+\frac1{d(d-2)n^2}
 \right)\,. \labell{broken}
\end{equation}
If $\imu$ increases beyond this bound (with fixed $n$), the event horizon
disappears and we are left with a naked singularity. Typical results for the
charged \ren entropy with imaginary chemical potential are shown in figure
\ref{imthree}. In comparing the figures, we see that while the charged \ren
entropy increases slowly with increasing $\mue$ in figures \ref{three} and
\ref{four}, $\tS_n(\imu)$ decreases, and in a much more dramatic fashion, as
$\imu$ increases in figure \ref{imthree}.


\subsection{Twist operators} \labell{pentagon}

In section \ref{PGtwist}, we derived various expressions for the conformal
weight and the magnetic response of the twist operators, as well as various
expansion coefficients appearing in these. Here we wish to examine these
properties of the twist operators in the boundary CFT dual to the
Einstein-Maxwell theory \reef{action}.

To begin, recall eq.~\reef{weight} for the conformal weight of the twist
operator,
\begin{equation}
h_n(\mue) = \frac{2 \pi n}{d-1}R^d\left(\mathcal{E}(T_0,\mue=0)
- \mathcal{E}(T_0/n,\mue)\right).\labell{Tweight}
\end{equation}
The details of the holographic calculation of the corresponding energy
densities are given in appendix \ref{background}. Then using
eq.~\reef{Edensity}, the desired conformal weight can be written as
 \beqa
h_n(\mue) &=& n \pi\, \frac{L^{d-1}}{\lp^{d-1}}\,\left(x_n^{d-2}(1-x_n^2)
-\frac{x_n^{d-2}(d-2)}{2(d-1)}\left(\frac{\mue \lstar}{2\pi
L}\right)^2\right)\,.
 \labell{dozer}
 \eeqa
Here we might note that this result can be expressed entirely in terms of
parameters in the boundary theory using eqs.~\reef{cv} and \reef{CTresult},
which show that ${L^{d-1}}/{\lp^{d-1}}\sim C_T$ and $\lstar^2
{L^{d-3}}/{\lp^{d-1}}\sim C_V$.

Next we would like to recover eqs.~\reef{interest1} and \reef{BIGA} for the
expansion coefficients of the above conformal weight. Hence given
eq.~\reef{dozer}, we evaluated the following two coefficients:
 \beqa
h_{10}=\partial_n h_n(\mue)|_{n=1,\mue=0}&=& \frac{2\pi}{d-1}\frac{L^{d-1}}{\lp^{d-1}} \,,
 \labell{dozer1}\\
h_{02}=\partial^2_\mue h_n(\mue)|_{n=1,\mue=0}&=& - \frac{(d-2)(2d-3)}{4
\pi(d-1)^2}\, \frac{\lstar^2\,L^{d-3}}{\lp^{d-1}}\,.
 \labell{dozer2}
 \eeqa
At this point, using eq.~\reef{CTresult}, we can substitute for $C_T$ in terms
of the ratio ${L^{d-1}}/{\lp^{d-1}}$ in eq.~\reef{interest1} and we recover
precisely eq.~\reef{dozer1}. Similarly, using eqs.~\reef{limit5} and
\reef{housefire} in eq.~\reef{BIGA}, we reproduce exactly the expression in
eq.~\reef{dozer2}. We also note that $h_{01}=\partial_\mue
h_n(\mue)|_{n=1,\mue=0}=0$ from eq.~\reef{dozer}, which is in agreement with
our general expectations in section \ref{PGtwist}.

Now we turn our attention to the magnetic response, which was given by
eq.~(\ref{parmk}),
\begin{equation}
k_n(\mue) = 2 \pi n R^{d-1} \rho(n,\mue) \,.
 \labell{curweight}
\end{equation}
Now in the holographic framework, we know that the charge density, \ie $\langle
J_\tau\rangle$ is determined by the normalizable component of the gauge field
\reef{gauge}, however, we leave its determination, including the precise
normalization, to appendix \ref{norm}. Substituting the holographic result
\reef{Qdensity} into eq.~(\ref{curweight}), we find the magnetic response in
our holographic model is
\begin{equation}
k_n(\mue) = \frac{d-2 }{2}\,\frac{\lstar^2\,L^{d-3}}{\lp^{d-1}}\,n x_n^{d-2}\mue
\,.
 \labell{snore}
\end{equation}
Again, this result can be expressed entirely in terms of parameters of the
boundary CFT using eq.~\reef{cv}. As above, we would like to examine the
expansion coefficients of the above magnetic response and hence we calculate
 \beqa
k_{01}=\partial_\mu k_n(\mue)|_{n=1,\mue=0} &=& \frac{d-2}{2}
\,\frac{\lstar^2\,L^{d-3}}{\lp^{d-1}}\,,
 \labell{snore1}\\
k_{11}=\partial_n\partial_\mue k_n(\mue)|_{n=1,\mue=0} &=& \frac{d-2}{2(d-1)}
\,\frac{\lstar^2\,L^{d-3}}{\lp^{d-1}}\,.
 \labell{snore2}
 \eeqa
These two coefficients can then be compared with eqs.~\reef{km01x} and
\reef{km11x} using the results in appendix \ref{two point}. As expected, the
holographic expression \reef{cv} of $C_V$ yields a precise agreement of
eq.~\reef{snore1} above with eqs.~\reef{km01x} and \reef{interest2} in section
\ref{PGtwist}. Similarly, eqs.~\reef{km11x} and \reef{snore2} match exactly
using eqs.~\reef{limit5} and \reef{housefire}.

\subsection{Thermodynamics, stability and phase transitions}
 \labell{stability}

It is natural to investigate the thermodynamical properties of the charged
hyperbolic black holes. Thermodynamical instability in some regions of phase
space could be responsible for interesting features in the \ren entropies. One
could also imagine that at low temperatures, a different geometry would be
prefered over the black hole phase, and the system would go through a
Hawking-Page phase transition as the temperature is increased. It turns out
none of these features occur for charged hyperbolic black hole in the grand
canonical ensemble. The Gibbs free energy $G=(M-M_c)-TS-\mu q$ was calculated
in \cite{thermodynamics}:
\begin{equation}
G=-\frac{V_\Sigma\,r_H^{d-2}}{2\,\lp^{d-1}}\left(1+\frac{(d-2)}{2(d-1)}\left(\frac{\mu\,\lstar}
{2\pi\,L}\right)^2+\frac{r_H^2}{L^2}-m_c(\mue)\right)
\end{equation}
where $m_c(\mue)$ is the critical mass at which the temperature vanishes, \ie
\begin{equation}
m_c=-2(d-1)r_c^{n-1}\left(1-\frac{(d-1)r_c^2}{(d-2)L^2}\right)
\ \ \mbox{with} \ \ \ \frac{r_c^2}{L^2}=\frac{d-2}{d}\left(1+\frac{(d-2)}{2(d-1)}\left(\frac{\mu\,\lstar}
{2\pi\,L}\right)^2\right)
\end{equation}
One can check that the Gibbs energy is always negative and equals zero when the
black hole is extremal, excluding any Hawking-Page phase transition. The specific heat was calculated in \cite{thermodynamics} as well:
\begin{equation}
C_\mu=T \left(\frac{\partial S}{\partial T} \right)_\mu=\frac{8\pi^2(d-1) V_\Sigma T r_H^{d}}{(d-2)\lp^{d-1}}
\left(1+\frac{d}{d-2}\frac{r_H^2}{L^2}+\frac{d-2}{d-1}\,\left(\frac{\mue\,\lstar}{2\pi L}\right)^2\right)^{-1}
\end{equation}
which is always positive, meaning the black holes are
thermodynamically stable.

Before leaving the subject of instabilities, it is interesting to note that the
presence of a light scalar in the bulk would render the black holes unstable at
low temperatures. Indeed, the extremal charged black hole has a $AdS_2\times
H^{d-1}$ near horizon geometry, where the relative radii of the two
spaces depends on the charge:
\begin{equation}
L^2_{AdS_2}=\frac{2L^2_{AdS_{d+1}}}{f''(r_H)} \ \ \ \ \ \ \ \ L^2_{{H}^{d-1}}=r_H^2
\end{equation}
with $f(r)$ is the metric function in eq.~\reef{funct} above. For simplicity,
let us consider a neutral scalar. The extremal black hole will develop scalar
hair if the mass of the scalar is below the BF bound of the near-horizon
$AdS_2$. We wish to consider normalizable modes and these must depend on the
hyperbolic coordinates, as the volume of the hyperboloid is infinite.
Normalizable modes can be expanded in eigenvalues of the Laplacian as
$\nabla_{H^{d-1}}^2\phi=-\lambda \phi$ with $\lambda>(d-2)^2/4$ and near the
horizon, this Laplacian will generate an effective shift of the mass of the
scalar. Thus we exepct an instability if the scalar mass $M$ lies in the range
\begin{equation}
-\frac{d^2}{4}<M^2L^2<-\frac{f''(r_H)}{8}-\frac{(d-2)^2}{4r_H^2}\ .
\end{equation}
It turns out the two terms on the right-hand side of this equation combine in
such a way that the answer does not depend on charge:
\begin{equation}
-\frac{d^2}{4}<M^2L^2<-\frac{d(d-1)}{4}
\end{equation}
found in \cite{renyiphases}. It seems that at zero temperature, a neutral
scalar will not detect changes in the geometry induced by charge. Turning our
attention to charged scalars, it was noted in \cite{holographicsuperconductor}
that the effect of the charge will be to induce a shift in the scalar mass.
This only makes things worse and a scalar instability is therefore expected as
well. We leave the detailed analysis of these effects for future work but note
that the \ren entropies should exhibit phase transitions as $n$ is varied, if
light scalars are present in the bulk spectrum \cite{Belin:2014mva}.

\section{Discussion} \labell{discuss}

We have examined a new class of entanglement measures \reef{charen} which extend the
usual definition of \ren entropy to include a chemical potential for a
conserved global charge. These charged \ren entropies measure the degree of
entanglement in different charge sectors of the theory. As described in section
\ref{two}, the evaluation of these entropies proceeds as usual with a Euclidean
path integral, but with the addition of a (fixed) background gauge field
which introduces a Wilson line, proportional to the chemical potential, around
the entangling surface. The latter can be interpreted as binding a sheet
of magnetic flux to the standard twist operators which appear in evaluating the
\ren entropy.

For the special case of a CFT with a spherical entangling surface, we can apply
a conformal transformation to map charged \ren entropies to the thermal
entropies of a grand canonical ensemble, albeit on the hyperbolic cylinder
$R\times H^{d-1}$. This allows us to study the properties of the generalized
twist operators, as discussed in section \ref{PGtwist}. In particular, the
conformal weight and the magnetic response of these twist operators are related
to the energy density and the charge density, respectively, in the thermal
ensemble on the hyperbolic cylinder. These two parameters are functions of both
$n$ and $\mue$ and exhibit certain universal characteristics when expanded in
the vicinity of $n=1$ and $\mue=0$.

In section \ref{Einstein}, we considered computations of charged \ren entropies
using holography, where they are related to the thermal entropy of charged
black holes with hyperbolic horizons. In addition to determining the charged \ren entropy, we
were able to determine the conformal weight and magnetic response of the
corresponding twist operators in this holographic model. In a particular, we
were able to recover the universal behaviour exhibited by the expansion
coefficients in, \eg eq.~\reef{interest2}. In section \ref{Gen twist op d2} and
appendix \ref{free}, we described the computation of charged \ren entropies for
free field theories. We found the results to be in qualitative agreement with
our holographic calculations. A particularly interesting point of comparison
is $d=2$, which was considered in section \ref{Gen twist op d2} for free
fermions, and in appendix \ref{threed} for holographic models.
For free bosons, we observed that the \ren entropy is non-analytic at $\imu=0$.
Thus, while there is a range where free fermions can be analytically continued to the real chemical potential,
 free bosons can not be so continued.

We found that the charged \ren entropy in the holographic model
 obeyed various inequalities \reef{inequalities}, which were
originally established for the standard \ren entropy without a chemical
potential. Following \cite{renyi}, we argued that the stability of the grand
canonical ensemble on the hyperbolic geometry was sufficient to guarantee these
inequalities would be satisfied by the charged \ren entropy. However, if one
examines the origin of these inequalities \cite{renyi1}, the derivation only
relied on the fact that the \ren entropies are moments of a probability
distribution with $p_i> 0$ and $\sum_i p_i = 1$. The same statement
 applies for the charged \ren entropies (with real chemical potential)
and so we can expect that eq.~\reef{inequalities} will be satisfied quite
broadly for these new entanglement measures. It would be interesting to
explicitly study the validity of these inequalities for more general QFT's and
choices of entangling surface. At the same time, it would be interesting to
investigate whether derivatives of $S_n(\mue)$ with respect to $\mue$ also
satisfy any general properties. For example, in figures \reef{three} and
\reef{four}, it seems that $\mue \frac{\partial S_n(\mue)}{\partial \mue}\ge0$
for our holographic model.
Note that an imaginary chemical potential does not
respect the above inequality.  In particular, the \ren entries and the free energies can take negative values.
The analytic continuation between the imaginary and the real chemical potentials is non-trivial because of poles and
branch cuts.
In gravity, the regular black hole space time ceases to exist  
for large $\imu$, \ie eq.~\reef{broken}.


There are several natural generalizations of the investigations presented here.
For example, the holographic computations could be extended to consider bulk
theories with higher derivative interactions (\eg Gauss-Bonnet or $F^4$ terms),
following \cite{GBBH1,GBBH2,anninos}. Another interesting direction would be to
connect our holographic calculations to the large-$N$ limit of super \ren
entropy for the ABJM model in \cite{susy}.

It may also be of interest to consider a generalization of the \ren entropy for
fixed charge ensembles (instead of fixing the chemical potential). Here, the
holographic computations may produce some interesting new behaviour. Finally,  in the case of a spherical entangling surface
(where the system is rotationally invariant) it is natural to label the states
by their angular momentum and introduce a conjugate chemical potential to
produce a `rotating \ren entropy' --- see also \cite{juan}. The
corresponding holographic calculations would involve more general classes of
spinning hyperbolic black holes.  A study of such rotating \ren entropies at
fixed angular potential, as well as charge, could follow very closely the
present discussion. Results along these lines will be presented in
\cite{spin2}.

\section*{Acknowledgments}
We would like to thank Horacio Casini, Steve Shenker and especially Tadashi
Takayanagi for helpful discussions. Research at Perimeter Institute is
supported by the Government of Canada through Industry Canada and by the
Province of Ontario through the Ministry of Research \& Innovation. AM and RCM
gratefully acknowledge support from NSERC Discovery grants. Research by RCM is
further supported by funding from the Canadian Institute for Advanced Research.
TS acknowledges support from an NSERC Graduate Fellowship.

\appendix

\section{Free QFT computations } \labell{free}

Here we consider various calculations of charged \ren entropies for free fields
using the heat kernel methods on hyperbolic spaces, and also by direct summing
of appropriate modes on spheres. These QFT computations are most readily done
if the chemical potential takes is purely imaginary values, \ie $\mue = i \imu$
where $\imu$ is real. In this case, the chemical potential produces to a
non-trivial boundary condition. As in section \ref{two}, we are interested in
conformal theories and hence we consider calculations for a massless
conformally coupled and complex scalar and for a free massless Dirac fermion.
In both cases, the global charge is simply related with phase rotations of the
corresponding field.

\subsection{Heat kernels on $S^1\times H^{d-1}$}
\label{Heat kernels on hyper}

To begin, we gather a few useful results for heat kernel methods
\cite{heatkernel}. First,  heat kernels on a product manifold factorize, for
both fermions and bosons,
 \be K_{S^1 \times H^{d-1}}(\{x_i\}, \{y_i\},t) =
K_{S^1}(x_1,y_1,t) K_{H^{d-1}}(x_{2,\cdots, d}, y_{2,\cdots,d},t)
 \ee
The total free energy in $S^1\times H^{d-1}$ is
 \be F = -\frac{(-)^f}{2} \int
\, d^dx \frac{dt}{t} e^{-m_c (1-f) t} K_{S^1\times H^{d-1}}(x,x,t)
 \ee
where $f=1$ for spin half Dirac fermions, and $f=0$ for scalars. The conformal
mass $m_c$ in $H^{d-1}$ for the conformally coupled scalar is \be m_c =
-\frac{(d-2)^2}{4R^2}, \ee where $R$ is the radius of $H^{d-1}$. For
convenience, we will set $R=1$ in the following.

We will consider finite temperature and purely imaginary chemical potential
$\mu_E = i \imu$ for a $U(1)$ global symmetry.
This can be incorporated into the heat kernel by setting appropriate boundary conditions.
For example, with inverse temperature $\beta = 2\pi n $ we have the boundary condition
\be
K_{S^1\times H^{d-1}} (\{x_1+ 2\pi n, \cdots, x_d\},\{y_1,\cdots, y_d\},t) = (-)^f e^{ i n \imu} K_{S^1\times H^{d-1}}(\{x\},\{y\},t).
\ee
Let us first focus on $K_{S^1}(x_1,y_1,t)$.
It is not hard to show that
\be \label{bc}
K_{\mathbb{R}^1}(x_1,y_1,t) = \frac{1}{\sqrt{4\pi t}} e^{-\frac{(x_1-y_1)^2}{4t}}.
\ee
Summing over images we find
\be
K_{S^1}(x_1,y_1,t) = \frac{1}{\sqrt{4\pi t}}  \sum_{m\in \mathbb{Z}} e^{\frac{-(y_1-x_1+ 2\pi  m)^2}{4t}}  e^{(-i n\imu -i\pi f)m},
\ee
which  satisfies the boundary condition  \ref{bc}.
Note that using this method, $n\imu <1$. The final result for $\imu n>1$ should be obtained from $\imu n<1$ by folding. We  therefore expect discontinuities in the free energies when $\imu n $ takes integer values.

In the case where $\imu=0$ we recover the usual heat kernel at finite temperature $\beta = 2\pi n$. The $S^1$ circle has radius $2\pi n$.

The heat kernel for massless scalars in $H^{D}$ takes the form \be
K^b_{H^{2x+1}}(\rho,t) = \frac{1}{(4\pi t)^{1/2}}\left( \frac{-1}{2\pi R^2
\sinh\rho} \frac{\partial}{\partial \rho}\right)^n e^{-x^2 t/R^2 - \rho^2
R^2/(4t)}, \ee for hyperbolic spaces of odd dimensions, and \be
K^b_{H^{2(x+1)}}(\rho,t) = e^{-(2x+1)^2t/(4R^2)}\left(\frac{-1}{2\pi R^2
\sinh\rho}\frac{\partial}{\partial \rho}\right)^n f_{H^2}(\rho,t), \ee where
$x$ is an integer, and $\rho$ is the geodesic distance between two points $x$
and $y$. The function $f_{H^2}(\rho,t)$ is defined as
 \be
f_{H^2}(\rho,t) = \frac{\sqrt{2}}{(4\pi t)^{3/2}}\int_\rho^\infty
d\tilde{\rho}\frac{\tilde{\rho}
e^{-\tilde{\rho}^2/(4t)}}{\sqrt{\cosh\tilde{\rho} -\cosh\rho  }}
 \ee
For fermions, we have
 \be
K^f_{H^{2x+1}}(\rho,t) = U(x,y)
\cosh(\rho/2)\bigg(\frac{-1}{2\pi}\frac{\partial}{\partial \cosh\rho}\bigg)^n
\cosh(\rho/2)^{-1} \frac{e^{-\rho^2/(4t)}}{\sqrt{4\pi t}}
 \ee
 and
 \be
K^f_{H^{2x}}(\rho,t) = U(x,y)
\cosh(\rho/2)\bigg(\frac{-1}{2\pi}\frac{\partial}{\partial \cosh\rho}\bigg)^n
\cosh(\rho/2)^{-1} k_{H^2}(\rho,t)
 \ee
and
  \be
K_{H^2}(\rho,t)=  \frac{\sqrt{2}\cosh^{-1}(\rho/2)}{(4\pi
t)^{3/2}}\int_\rho^\infty d\tilde{\rho}\frac{\tilde{\rho}
\cosh{\tilde{\rho}/2}e^{-\tilde{\rho}^2/(4t)}}{\sqrt{\cosh\tilde{\rho}
-\cosh\rho  }}
 \ee
The matrix $U(x,y)$  has a trace given by $2^{[d/2]}$, where $[\cdots]$ denotes
the integer part.  It counts the dimension of spinor space in $d$ dimensions. 

From eq.~\reef{ren3}, we can write the charged \ren entropy as
 \be
S_n = \frac{F(n,\imu) - n F(1,\imu)}{n-1}
 \ee
where $F(n,\imu)\equiv-\log Z_n(\imu)$ is the free energy evaluated at
temperature $\beta = 2\pi n$ and chemical potential $\imu$. We are thus ready
to compute free energies at different dimensions.

\subsubsection{$d=2$}

At temperature $\beta_n = 2\pi n$ and finite chemical potential $\imu$, the
free energy is
 \be
F(n,\imu) = \frac{(-)^f}{2}(2\pi n) V_{H^{d-1}} \int \, \frac{dt}{t} \sum_m
\frac{e^{-\pi^2m^2/t}}{\sqrt{4\pi t}} e^{(-i n\imu -i\pi f)m}
K_{H^{1}}(\rho=0,t).
 \ee
The heat kernel is
\be
K_{H^1}(0,t) = K_{\mathbb{R}}(0,t) = \frac{1}{\sqrt{4\pi t}}.
\ee
The free energy $F(n,\imu)$ is divergent due to the $m=0$ mode in the $S^1$ heat kernel. For $m=0$ the contribution is linear in $n$, where $n$ appears as the overall volume factor from $S^1$.  This dependence therefore drops out from $S_n$. We could therefore rewrite the regulated free energy  $\hat{F}(n,\imu)$ as
\ba
\hat{F}(n,\imu) &&= \frac{(-)^f}{2}(2\pi n) V_{H^{d-1}} \int \, \frac{dt}{t} \sum_{m \in \mathbb{Z}_+} \frac{e^{-\pi^2m^2/t}}{\sqrt{4\pi t}} 2\cos({  (n\imu + \pi f)m}) K_{H^{1}}(0,t)  \nonumber \\
&&=\frac{(-)^f}{2}V_{H^{d-1}} \sum_{m\in \mathbb{Z}_+}  \frac{ 8  \cos(
   n \, m \imu ) \cos(m \pi f)}{8\pi^2 n\, m^2} \nonumber \\
&&= \frac{(-)^f}{4\pi^2 n}V_{H^{d-1}} \bigg(\Li_2(e^{ i n \imu+i\pi f})+ \Li_2(e^{ -i n \imu-i\pi f})  \bigg)
\eea
For $0\le$Re$(x)<1$ and Im($x)\ge 0$, or Re$(x)\ge 1$ and Im$(x)<0$
\be
\Li_2(e^{2\pi i x}) +\Li_2(e^{-2\pi i x}) = -\frac{(2\pi i)^2}{2} B_2(x) = - \frac{(2\pi)^2}{\Gamma[2]} \zeta(-1, x),
\ee
where $B_2$ is the Bernoulli polynomial, and $\zeta(a,b)$ the Hurwitz zeta function.
We are left with
\be
\hat{F}(n,\imu) = \frac{(-)^{f} V_{H^1}}{2n} B_2\left(\frac{n\,\imu}{2\pi} + \frac{f}{2}\right) = \frac{(-)^{f} V_{H^1}}{2n} \bigg(
\frac{1}{12} (2 - 6 f + 3 f^2) + \frac{(f-1 ) n \imu}{2\pi} + \frac{n^2 \imu^2}{4\pi^2}\bigg).
\ee

For fermions, the linear term in $\imu$ vanishes, as expected. However, for bosons there is a linear $\imu$ term despite the fact that the sum is explicitly even. This term appears from a term
$n\imu \ln(n \imu) - n\imu \ln(-n\imu)$ in the expansion of the poly-log in $\imu$. This suggests that we are actually taking the absolute value of the linear term. This can be readily confirmed by computing the sum numerically. As a result, the free energy has a diverging slope at $\imu=0$, suggesting a phase
transition there. There are also phase transitions whenever $\imu n$ is an integer, as noted above. At precisely $\imu n = 1/2$, the
first derivative w.r.t. $\imu$ jumps from zero to $V_{H^1}$.

The $\imu^2$ term cancels out in the \ren entropy (since it is linear in $n$) for $\imu n <1/2$.
The result for a Dirac fermion is
\be
S^f_n = \frac{c}{6}(1+\frac{1}{n}) V_{H^1}
\ee
Note that $c=1/2$ for a single Majorana fermion, but this should be doubled for a charged fermion.
This reproduces the result obtained via the twist operator method  in the main text.

For bosons we obtain instead
\be
S^b_n = c \left(\frac{1}{6}(1+ \frac{1}{n}) - \frac{|\imu|}{2\pi} \right)V_{H^1},
\ee
Again, $c=1$ for a real boson, which must be doubled for a charged boson.
One might worry that the result for bosons does not appear to agree with that of fermions given that they are
related by bosonization in 1+1 dimensions. We note however that via bosonization of $U(1)$ charged fermions,
the corresponding bosons transform by translation, and thus should instead satisfy the following boundary condition
:$\phi(\tau + 2\pi) =  \phi(\tau) + n \imu$.
Therefore, our computation for charged bosons is not related to charged fermions by bosonization.
Another point to note is that with the absolute sign, the bosonic result is not analytic even for arbitrarily small
$\imu$, such that it does not analytically continue to the complex plane, as in the case for fermions.

\subsubsection{$d=4$}
Let us work out one more example where there is non-trivial $\imu$ dependence.
At $d=3+1$, the main difference is the heat kernel for both bosons and fermions on $H^3$.
For bosons, the equal-point heat kernel is
\be
e^{-m_s t} K^b_{H^3}(0,t) = \frac{1}{(4\pi t)^{3/2}} ,
\ee
where we have substituted in the conformal mass of the scalar.
For a Dirac fermion, the heat kernel is
\be
K^f_{H^3}(0,t) = 4 \frac{1}{(4\pi t)^{3/2}} (1 + \frac{t}{2})
\ee
Following the same steps as in $d=3+1$, and again ignoring the $m=0$ term, the
\ren entropy becomes, for bosons :
\ba
S_n &&= \tr(1) V_{H^3}\sum_{m\in \mathbb{Z}_+} \frac{n^4 \cos( m  \imu) -
 \cos( m  n \imu)}{8 m^4 \pi^5 (n-1) n^3}  \nonumber \\
&&= -\frac{1}{8\pi^5 (n-1) n^3 } \bigg(-n^4 (\Li_4( e^{- i  \imu} )+
     \Li_4(e^{ i \imu}) ) + \nonumber \\
&&\Li_4( e^{- i n \imu}) + \Li_4(e^{ i n \imu})\bigg)
\eea
where $\tr(1) = 2$ for a pair of real bosons (which together form a complex $U(1)$ charged boson ).
The above combinations of poly-logs again admit a representation in terms of the Bernoulli polynomial. Altogether we have
\be
S^b_n = \frac{V_{H^3}}{2\pi}\bigg(\frac{1 + n + n^2 + n^3}{180 n^3} - \frac{(n+1) \imu^2}{24\pi^2 n} + \frac{\imu^3}{24\pi^3} \bigg)
\ee
Again we are left with a $\imu^3$ term that is odd in $\imu$, and we argue that this term should be enclosed inside an absolute sign since our summation is even. As a result, once again we lose analyticity even for arbitrarily small values of $\imu$.

Now let us also look at the corresponding result for fermions.
The \ren entropy is
\ba
S^f_n &&= V_{H^3}\sum_{m\in \mathbb{Z}_+}\frac{(-1)^m (2 + m^2 \pi^2 n^2) \cos( m n \imu)}{m^4 4\pi^5 n^3} \nonumber \\
&&= V_{H^3}\frac{(1 + n) (7 + n^2 (37 - 120 \imu^2)}{1440\pi n^3},
\eea
 which interestingly, is again automatically even in $\imu$, and that for purely imaginary $\imu$, is positive definite.
We note that the above expression is not positive definite in $\imu$. We find that for sufficiently large value of $\imu$ while within the interval $n \imu <1/2$ that the above expression can turn negative. This is as expected since the trace
\be
\tr\rho^n= \tr(e^{-n ( H  - i \imu Q)})
\ee
 is not necessarily positive definite quantity. When $\imu$ is purely imaginary, we return to the usual thermodynamic chemical potential and the trace should be positive definite.
Note that the \ren entropy for the fermions, which admit analytic continuation for small values of $\imu$,
is indeed positive definite when $\imu$ is purely imaginary.

{\flushleft \bf{Remark: $d=2+1$.}}
Here the complication is the more complicated form of the heat kernel in $H^2$. Because of that, it doesn't have a neat analytic result, but one can evaluate these results numerically. We find precise agreement with the calculation obtained on a sphere in later sections, and we will not repeat the details here.

\subsection{Wavefunctions on $S^1\times H^{d-1}$}
We can reproduce the heat kernel results by analyzing the wave functions on the hyperbolic space.
This method was used in \cite{subir} to study the R\'{e}nyi entropy of the free theories without the chemical potential.
In this subsection, we generalize \cite{subir} to include the chemical potential.

\subsubsection{Free scalar field}

We consider a free boson on a $S^1\times H^{d-1}$ with $H^{d-1}$ radius $R$
\ba
S=\int d^{d}r\sqrt{g}(|\pd_{\mu}\phi|^2+M^2|\phi|^2)
\eea
where $M$ is the conformal mass.
The metric is
\ba
ds^2=d\theta^2+R^2(d\eta^2+\sinh^2\eta d\Omega_{d-2}^2).
\eea
The periodicity of the $S^1$ time circle $(\theta)$ is $2\pi n$.
The Wilson loop changes boundary condition around the time circle from
$\phi(2\pi n)=\phi(0)$ to
\ba
\phi(2\pi n)=e^{i n\imu}\phi(0).
\eea
The mode function satisfying this boundary condition is
\ba
e^{i\left({m\over n}+\frac{\imu}{2\pi} \right)\theta},
\eea
where $m$ is an integer.
The eigenvalue of the Laplace operator $-\Delta-M^2$ is
\ba
\lambda+\left({m\over n}+\frac{\imu}{2\pi}\right)^2,~~~~~\lambda\ge 0.
\eea
We define the free energy
\ba
F_n&=&\Tr \log\left(-\Delta-M^2 \right) \cr
&=&\sum_{m\in \mathbb{Z}}\int_{0}^{\infty}d\lambda \mathcal{D}(\lambda) \log
\left(
\lambda+\left({m\over n}+\frac{\imu}{2\pi}\right)^2
\right)\\
&=&\int_{0}^{\infty}d\lambda \mathcal{D}(\lambda)
\left(
2\pi n \sqrt{\lambda}+
\log
\left(1-2\cos( n\imu)e^{-2\pi n \sqrt{\lambda}}+e^{-4\pi n \sqrt{\lambda}}
\right)
\right)
\label{boson free energy}
\eea
where $\mathcal{D}(\lambda)$ is the density of states.
In the last equation, we used the following formula for the regularized sum
 \ba
\sum_{k\in \mathbb{Z}} \log\left({(k+\alpha)^2\over n^2}+a^2
\right)=\log[2\cosh(2\pi n |a|) -2\cos(2\pi \alpha) ].
 \labell{log-sum}
 \eea
In the case of scalar bosons,
the density of states $\mathcal{D}(\lambda)$ on $H^{d-1}$ is given by \cite{density-of-states}
\ba
\mathcal{D}(\lambda)d\lambda={\text{vol}(H^{d-1})\over (4\pi)^{d-1\over 2}\Gamma\left({d-1\over 2}\right)}
{|\Gamma(i\sqrt{\lambda}+{d-2\over 2})|^2\over \sqrt{\lambda} |\Gamma(i\sqrt{\lambda})|^2} d\lambda.
\eea
The explicit forms for low dimensions are
\ba
d-1=1;&&\mathcal{D}(\lambda)d\lambda={\text{vol}(H^1) \over 2\pi  \sqrt{\lambda}}d\lambda \cr
d-1=2;&&\mathcal{D}(\lambda)d\lambda= {\text{vol}(H^2)\over 4\pi}\tanh(\pi\sqrt{\lambda})d\lambda \cr
d-1=3;&&\mathcal{D}(\lambda)d\lambda={\text{vol}(H^3) \over (2\pi)^2}\sqrt{\lambda} d\lambda
\eea
$\text{vol}(H)$ is the regularized volume of the hyperbolic space.

The first term in (\ref{boson free energy}) is divergent and needs a regularization. However, it will not contribute to the R\'{e}nyi entropy since it linearly depends on $n$.
One can show that this integration reproduces the heat kernel results.

\subsubsection{Free Dirac fermion  \label{2dfermion}}
We also consider a free Fermion
\ba
S=\int d^{d}x \sqrt{g}\bar{\psi}(i\slashed{D})\psi,
\eea
The free energy is
\ba
F_{n}=-\Tr \log (i\slashed{D}),
\eea

In the presence of the Wilson loop, the boundary condition of $\psi$ along the time circle changes
from $\psi(2\pi n)=-\psi(0)$ to
\ba
\psi(2\pi n)=-e^{i n \imu }\psi(0)
\label{QFT bc with holonomy}
\eea
So the eigenfunction along $\theta$ is $e^{im \theta/n}$ with
\ba
m \in \mathbb{Z}+{1\over 2}+\frac{n \imu}{2\pi} .
\eea
The eigenvalue spectrum of $(i\slashed{D})$ is
\ba
\pm \sqrt{\lambda^2+{m^2\over n^2}}.
\eea
The free energy is
\ba
F_{n}&=&-{1\over 2}\sum_{m\in \mathbb{Z}+{1\over 2}}\int_{0}^{\infty}d\lambda \mathcal{D}(\lambda)
\log\left(
\lambda^2+{m^2\over n^2}
\right)\cr
&=&-{1\over 2}\int_{0}^{\infty}d\lambda
\mathcal{D}(\lambda)\log
\left(
2\cosh(2\pi n \lambda)+2\cos( n \imu)
\right)
\label{fermion-free-energy}
\eea
As before, we used (\ref{log-sum}) in the last equation.

The density of states $\mathcal{D}(\lambda)$  in $d-1$ dimensions is \cite{density-of-states}
 \be
{\mathcal{D}(\lambda)\over \text{vol}(H^{d-1})}=\left({\Gamma\left({d-1\over 2}\right)2^{d-4}\over \pi^{(d-1)/2+1}}2^{\left[{(d-1)\over 2}\right]}\right)
{2^{4-2(d-1)}\over (\Gamma((d-1)/2))^2}\cosh(\pi \lambda)
\left|
\Gamma \left(i\lambda +{(d-1)\over 2}\right)
\right|^2.
 \ee
Here, $\mathcal{D}(\lambda)$ is normalized so that the spinor $\zeta$-function per unit volume is given by
 \ba
\text{tr}(-\slashed{D}^2+m^2)^{-s}=\int_{0}^{\infty}(\lambda^2+m^2)^{-s}{\mathcal{D}(\lambda)\over \text{vol}(H^{d-1})}d\lambda.
 \eea
For odd $d-1$, it is
 \ba
{\mathcal{D}(\lambda)\over \text{vol}(H^{d-1})}={\pi\over 2^{2(d-3)}(\Gamma((d-1)/2))^2}
\prod^{(d-3)/2}_{j={1\over 2}}(\lambda^2+j^2)
 \eea
and for even $d-1$
 \ba
{\mathcal{D}(\lambda)\over \text{vol}(H^{d-1})}={\pi \lambda \coth(\pi\lambda)\over 2^{2(d-3)}(\Gamma((d-1)/2))^2}
\prod^{(d-3)/2}_{j=1}(\lambda^2+j^2)
 \eea
The explicit forms for low dimensions are
 \ba
d-1=1 ;&& \mathcal{D}(\lambda)d\lambda= {\text{vol}(H^1)\over \pi}d\lambda\cr
d-1=2 ;&&\mathcal{D}(\lambda)d\lambda={\text{vol}(H^2)\over \pi}\lambda
\coth(\pi \lambda)d\lambda\cr
d-1=3;&&\mathcal{D}(\lambda)d\lambda=\text{vol}(H^3) \left(\lambda^2+{1\over 4}\right)d\lambda
 \eea

The first term in (\ref{fermion-free-energy}) diverges and needs a
regularization, while the second term is finite. We can regularize the
divergence using zeta function regularization or flat space
subtraction. In any case, it doesn't contribute to the R\'{e}nyi entropy since
it is linear in $n$.
The final result agrees with the twist operator computation (Sec \ref{Gen twist op d2}), the heat kernel computation (Appendix \ref{Heat kernels on hyper})
and the wave function computation on a sphere (Appendix \ref{wavefunc on s})

%
%
%

\subsection{Wavefunctions on $S^d$}
\label{wavefunc on s}
Another convenient way of computing the R\'{e}nyi entropy of CFT is to map onto a sphere.
Let us consider a scalar field on $S^3$.
The metric is
\ba
ds^2=\cos^2\theta d\tau^2+d\theta^2+\sin^2\theta d\phi^2
\eea
with $0\le \tau < 2\pi n, 0\le \phi<2\pi,$ and $0\le \theta <\pi/2$.
Because of the periodicity of $\tau$, there is a conical singularity at $\cos\theta=0$.
We can do the heat kernel analysis on the sphere in a similar way to the hyperbolic case. However, we need to set a regularity condition at the conical singularity.
An alternative way to compute the free energy is to look at the wave functions and their eigenvalues directly.
The analysis below is a generalization of \cite{subir} and we cite several results from their paper.

\subsubsection{Free scalar field}
The free energy of the free scalar field on the sphere is
\ba
F_{n}=\text{tr}\log \left(-\Delta+{\mathcal{R}\over 8}\right)
\label{boson free e}
\eea
where $\mathcal{R}=6$ for $3d$ case.
We assume that the eigenfunction of the Laplacian takes the form  $f(\theta)e^{im_{\tau}\tau+im_{\phi}\phi}$.
The function $f(\theta)$ obeys the following equation
\ba
f''(\theta)+2\cot \theta f'(\theta)-\left({m^2_{\tau}\over\cos^2\theta}+{m^2_{\phi}\over \sin^2\theta}\right)f(\theta)=\lambda f(\theta).
\eea
From the regularity of $f(\theta)$, the eigenvalue $\lambda$ is fixed to
\ba
\lambda=-s(s+2), ~~~~~~s=|m_{\tau}|+|m_{\phi}|+2a, ~~~~a\in \mathbb{N}
\eea
The periodicity of $\phi$ requires
$m_{\phi}$ to be quantized in $\mathbb{Z}$.
In the presence of the Wilson loop, the boundary condition of $\phi$ becomes
\ba
\phi(2\pi n)=e^{i n \imu }\phi(0)
\eea
Therefore, $m_{\tau}$ is quantized in ${\mathbb{Z}\over n}+\frac{\imu}{2\pi}$.

Let us denote
\ba
m_{\tau}={\alpha\over n}+\frac{\imu}{2\pi},~~~~~
m_{\phi}=\beta,~~~~~(\al,\beta\in\mathbb{Z})
\eea
and
\ba
\al=kn+p,~~~~~k\in\mathbb{Z},~~p\in [0,n-1].
\eea
The free energy (\ref{boson free e}) is
\ba
F_{n}=\sum_{k=-\infty}^{\infty}\sum_{p=0}^{n-1}\sum_{\beta=-\infty}^{\infty}\sum_{a=0}^{\infty}
\log\left(s(s+2)+{3\over4}\right)
\label{free energy formal}
\eea
with
\ba
s=\left|k+{p\over n}+\frac{\imu}{2\pi}\right|+|\beta|+2a.
\eea
We want to count the degeneracied for a given value of $s$.
Let us first assume that $\imu$ satisfies
$0\le \frac{\imu}{2\pi}<{1\over q}$.
Then the degeneracy for
\bea
s=m+{p\over n}+\frac{\imu}{2\pi},~~~m-{p\over n}-\frac{\imu}{2\pi}+1~~~(p\in[0,n-1],~~m\in \mathbb{N}),
\eea
is
\bea
{(m+1)(m+2)\over 2}.
\eea
Therefore, the free energy is
\bea
F_n&=&\sum_{p=0}^{q}\sum_{m=0}^{\infty}\Big[{(m+1)(m+2)\over 2}\Big\{
\log((m+{p\over q}+\frac{\imu}{2\pi}+1)^2-{1\over4})
+\log((m-{p\over q}-\frac{\imu}{2\pi}+2)^2-{1\over4})
\Big\}\Big]\cr
&=&\sum_{p=0}^{q}\sum_{a=a_1}^{a_4}\left[-{1\over 2}\left(\zeta'(-2,a)+(3-2a)\zeta'(-1,a)+(a^2-3a+2)\zeta'(0,a)\right)\right]
\label{free energy near zero}
\eea
where $\zeta'(s,a)={\pd \zeta(s,a)\over \pd s}$ and
\be
a_1=({p\over n}+\frac{\imu}{2\pi}+{1\over 2}),~~a_2=({p\over n}+\frac{\imu}{2\pi}+{3\over 2}),
~~a_3=(-{p\over n}-\frac{\imu}{2\pi}+{3\over 2}),~~a_4=(-{p\over n}-\frac{\imu}{2\pi}+{5\over 2}).
\ee
The expansions near $\imu=0$ are
\bea
F_1&=&{\log2\over 4}-{3\zeta(3)\over 8\pi^2}\cr
&&
-\frac{\imu ^2 }{96\pi^2} \left(2 \left(-12
   \zeta^{(1,1)}\left(-1,\frac{1}{2}\right)-24
   \zeta^{(1,1)}\left(-1,\frac{3}{2}\right)-12
   \zeta^{(1,1)}\left(-1,\frac{5}{2}\right)+3
   \zeta^{(1,2)}\left(-2,\frac{1}{2}\right)\right.\right.\cr
 &&  +6
   \zeta^{(1,2)}\left(-2,\frac{3}{2}\right)+3
   \zeta^{(1,2)}\left(-2,\frac{5}{2}\right)+6
   \zeta^{(1,2)}\left(-1,\frac{1}{2}\right)-6
   \zeta^{(1,2)}\left(-1,\frac{5}{2}\right)\cr
 &&\left. \left.+28-36 \log 2+6\log
   3\right)+3 \pi ^2\right)+\mathcal{O}(\imu^4),
   \eea
\bea
F_2&=&{\log2\over 4}+{\zeta(3)\over 8\pi^2}\cr
&&+{\frac{\imu^2}{4\pi^2}}\left(
\frac{1}{12} \left(12 \zeta^{(1,1)}\left(-1,\frac{1}{2}\right)+24
   \zeta^{(1,1)}(-1,1)+24
   \zeta^{(1,1)}\left(-1,\frac{3}{2}\right)+24
   \zeta^{(1,1)}(-1,2)+\right.\right.\cr
&&   12  \zeta^{(1,1)}\left(-1,\frac{5}{2}\right)-3
  \zeta^{(1,2)}\left(-2,\frac{1}{2}\right)-6
   \zeta^{(1,2)}(-2,1)-6
   \zeta^{(1,2)}\left(-2,\frac{3}{2}\right)\cr
&&   -6
   \zeta^{(1,2)}(-2,2)-3
   \zeta^{(1,2)}\left(-2,\frac{5}{2}\right)-6
   \zeta^{(1,2)}\left(-1,\frac{1}{2}\right)-6
   \zeta^{(1,2)}(-1,1)+6 \zeta^{(1,2)}(-1,2)\cr
&& \left. \left.    +6
   \zeta^{(1,2)}\left(-1,\frac{5}{2}\right)-40-6 \log (3)+12 \log (16 \pi
   )\right)-\frac{\pi ^2}{8}
\right)+\mathcal{O}(\imu^4).
\eea
Each of the functions $\zeta^{(1,1)}, \zeta^{(1,2)}$ etc has some subtlety in evaluation.
We may always go back to the expression (\ref{free energy near zero}) when we evaluate the free energy explicitly.
The leading terms agrees with the results in \cite{subir}.

\begin{figure}[h!]
\includegraphics[width=\textwidth]{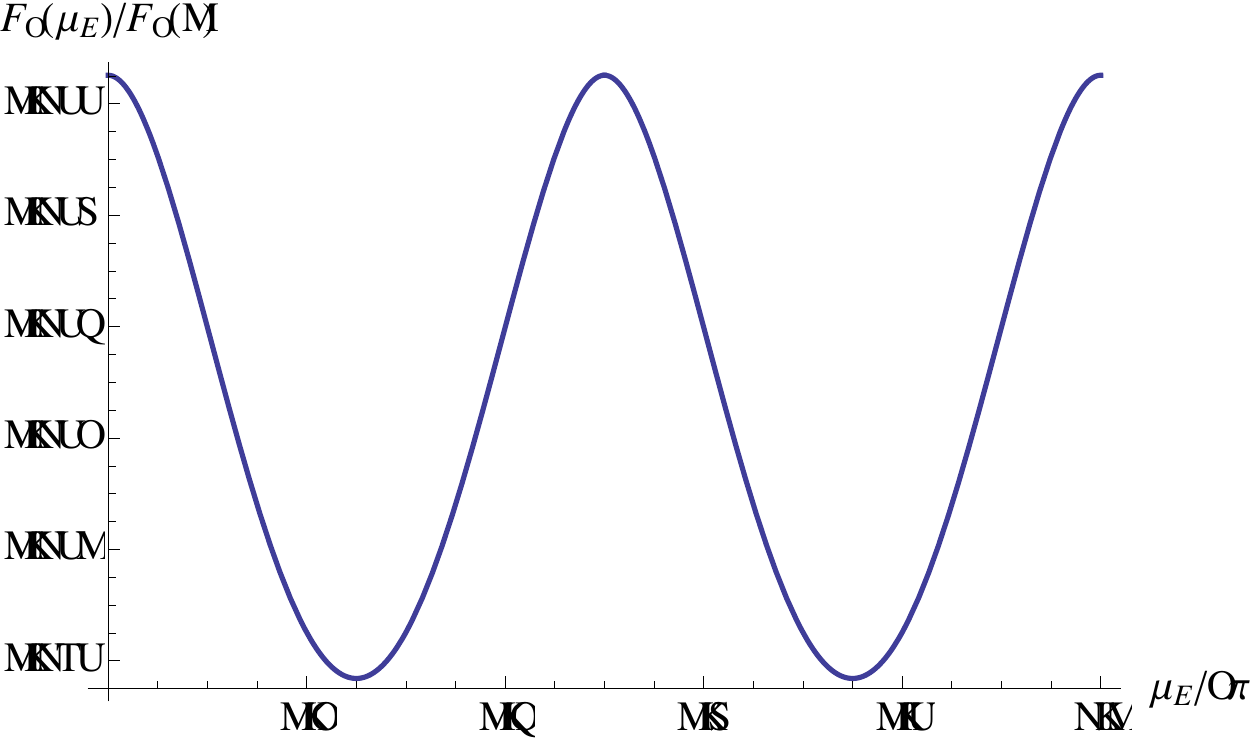}
\caption{Boson free energy $n=2$}
\label{free boson plot}
\end{figure}

The expression (\ref{free energy near zero}) is not valid for $\frac{\imu}{2\pi}>{1\over n}$.
In this case, there is a number $p_1\le n-1$ satisfying
\bea
-1+{p_1\over n}+\frac{\imu}{2\pi}<0,~~~~
-1+{p_1+1\over n}+\frac{\imu}{2\pi}<0.
\eea
By using this number, the eigenvalues and their degeneracies become
\bea
{(m+1)(m+2)\over 2}&~~~~\text{for}~~&s=m+{p\over n}+\frac{\imu}{2\pi},~~m-{p\over n}-\frac{\imu}{2\pi}+2~~(p\in[0,n-1]),\cr
(m+1)&~~~~\text{for}~~&s=m-{p\over n}-\frac{\imu}{2\pi}+1 ~~(p\in[0,p_1]),~ \cr
(m+1)&~~~~\text{for}~~&s=m+{p\over n}+\frac{\imu}{2\pi}-1~~(p\in[p_1+1,n-1]).
\eea
The free energy is
\bea
F_n&=&\sum_{p=0}^{n-1}\sum_{m=0}^{\infty}{(m+1)(m+2)\over 2}\Big(
\log((m+{p\over n}+\frac{\imu}{2\pi}+1)^2-{1\over4})
+\log((m-{p\over n}-\frac{\imu}{2\pi}+3)^2-{1\over4})\Big)\cr
&&+\sum_{p=0}^{p_1}\sum_{m=0}^{\infty}{(m+1)}\Big(
\log((m-{p\over n}-\frac{\imu}{2\pi}+2)^2-{1\over4})\Big)\cr
&&+\sum_{p=p_1+1}^{n-1}\sum_{m=0}^{\infty}{(m+1)}\Big(
\log((m+{p\over n}+\frac{\imu}{2\pi})^2-{1\over4})\Big)\cr
&=&
\sum_{p=0}^{n}\sum_{a=b_1}^{b_4}(-{1\over 2}(\zeta'(-2,a)+(3-2a)\zeta'(-1,a)+(a^2-3a+2)\zeta'(0,a)))\cr
&+&\sum_{p=0}^{p_1}\sum_{a=c_1}^{c_2}(-(\zeta(-1,a)+(1-a)\zeta(0,a)))
+\sum_{p=p_1+1}^{n-1}\sum_{a=c_3}^{c_4}(-(\zeta(-1,a)+(1-a)\zeta(0,a)))
\eea
where
\be
b_1=({p\over q}+\frac{\imu}{2\pi}+{1\over 2}),~~b_2=({p\over q}+\frac{\imu}{2\pi}+{3\over 2}),
~~b_3=(-{p\over q}-\frac{\imu}{2\pi}+{5\over 2}),~~b_4=(-{p\over q}-\frac{\imu}{2\pi}+{7\over 2}),
\ee
and
\be
c_1=(-{p\over q}-\frac{\imu}{2\pi}+{3\over 2}),~~c_2=(-{p\over q}-\frac{\imu}{2\pi}+{5\over 2}),
~~c_3=({p\over q}+\frac{\imu}{2\pi}-{1\over 2}),~~c_4=({p\over q}+\frac{\imu}{2\pi}+{1\over 2}).
\ee
We show the numerical result in Fig.\ref{free boson plot}. It is remarkable that the function is smooth around $\imu={1\over n}$.

%
%
%

\subsubsection{Free Dirac fermion}

Next, we consider a spinor $\psi$ on $S^3$.
It satisfies the Dirac equation
\bea
i\sigma^{\mu}_a\gamma^a\left(\pd_{\mu}\psi+{1\over 4}\omega^{ab}_{\mu}\gamma_{ab}\psi\right)=\lambda \psi.
\eea
The periodic part of the spinor $\psi$ in $\tau$ and $\phi$ directions can be written as
$e^{im_{\tau}\tau+im_{\phi}\phi}$.
As shown in \cite{subir}, the regularity condition at $\theta=0$
restricts the allowed eigenvalue $\lambda$.
There are four types of eigenvalues : \\
Two positive $\lambda$
\bea
\text{Case 1}&&\lambda= m_{\tau}+m_{\phi}+{3\over2}+2a,  ~~~~~m_{\tau}\ge 0,~~~~~ m_{\phi}\ge -{1\over2}-a,~~ a\in \mathbb{N}, \cr
\text{Case 2}&&\lambda= -m_{\tau}+m_{\phi}+{1\over2}+2a,  ~~~~~m_{\tau}< 0,~~~~~ m_{\phi}\ge {1\over2}-a,~~ a\in \mathbb{N},
\eea
and two negative $\lambda$
\bea
\text{Case 3}&&\lambda= -(m_{\tau}+m_{\phi}+{1\over2}+2a),  ~~~m_{\tau}\ge 0,~~~ m_{\phi}\ge {1\over2}-a,~~ a\in \mathbb{N}, \cr
\text{Case 4}&&\lambda= -(-m_{\tau}+m_{\phi}+{3\over2}+2a),  ~~~m_{\tau}< 0,~~~ m_{\phi}\ge -{1\over2}-a,~~ a\in \mathbb{N},
\eea
where
\bea
m_{\tau} \in {1\over q}(\mathbb{Z}+{1\over 2}+\frac{q\imu}{2\pi}),~~~~~m_{\phi} \in \mathbb{Z}+{1\over2}
\eea

As before, we first consider the case $0\le\frac{n\imu}{2\pi} <{1\over 2}$.
In this case, the eigenvalues and the degeneracies are
\be
{(k+1)(k+2)\over 2} ~~~\text{for}~~~\lambda=\pm(k+{p\over q}+{1\over 2q}+\frac{\imu}{2\pi}+1),
~\pm(k+{p\over q}+{1\over 2q}-\frac{\imu}{2\pi}+1)
\ee
where $k\in \mathbb{N}$.

The free energy is
\bea
F_{n}&=&-2\sum_{p=0}^{q-1}\sum_{k=0}^{\infty}{(k+1)(k+2)\over 2}\log (k+1+{p\over n}+{1\over 2n}+\frac{\imu}{2\pi})\cr
&&-2\sum_{p=0}^{n-1}\sum_{k=0}^{\infty}{(k+1)(k+2)\over 2}\log (k+1+{p\over n}+{1\over 2n}-\frac{\imu}{2\pi})\cr
&=&
\sum_{p=0}^{q-1}\sum_{a=a_1}^{a_2}(\zeta'(-2,a)+
(1-2a)\zeta'(-1,a)
+a(a-1) \zeta'(0,a))
\label{free energy fermion sphere 3d}
\eea
where
\bea
a_1=1+{p\over n}+{1\over 2n}+\frac{\imu}{2\pi},~~a_1=1+{p\over n}+{1\over 2n}-\frac{\imu}{2\pi}
\eea
We show the explicit form of the free energy near $\imu=0$.
\bea
F_{1}&=&{\log 2\over 4}+{3\zeta(3)\over 8\pi^2}\cr
&+&\left(
-\log8-4\zeta^{(1,1)}(-1,{3\over 2})+\zeta^{(1,2)}(-2,{3\over 2})-{1\over 4}\zeta^{(1,2)}(0,{3\over 2})
\right)\frac{\imu^2}{4\pi^2}+\mathcal{O}(\imu^4) \cr
F_{2}&=&{3\zeta(3)\over 32\pi^2}+{3\log 2\over 16}+{G\over 2\pi}\cr
&+&\Big(
-9\log2+2\log3-4\zeta^{(1,1)}(-1,{5\over 4})-4\zeta^{(1,1)}(-1,{7\over 4})
-\zeta^{(1,1)}(0,{5\over 4})+\zeta^{(1,1)}(0,{7\over 4})\cr
&&+\zeta^{(1,2)}(-2,{5\over 4})+\zeta^{(1,2)}(-2,{7\over 4})
+{1\over 2}\zeta^{(1,2)}(-1,{5\over 4})-{1\over 2}\zeta^{(1,2)}(-1,{7\over 4})\cr
&&-{3\over 16}\zeta^{(1,2)}(0,{5\over 4})
-{3\over 16}\zeta^{(1,2)}(0,{7\over 4})
\Big)\frac{\imu^2}{4\pi^2}+\mathcal{O}(\imu^4)
\eea
where $G$ is the Catalan constant.
Again, the functions $\zeta^{(1,1)},\zeta^{(1,2)}$ etc are a formal expression and one may use (\ref{free energy fermion sphere 3d})
to evaluate the free energy.
The leading terms agree with \cite{subir}.
Notice that only even powers of $\imu$ appears in the expansion.
The expression (\ref{free energy fermion sphere 3d}) is not valid for $\imu>{1\over 2n}$.
In this region there is a number $p_1$ which satisfies
\bea
-1+{p_1\over n}+{1\over 2n}+\frac{\imu}{2\pi}<0,~~~-1+{p_1+1\over n}+{1\over 2n}+\frac{\imu}{2\pi}\ge0.
\eea
Then the eigenvalues, their degeneracies, and the range of $p$ are as follows:
\bea
{(k+1)(k+2)\over 2} ~~&&\text{for}~~\lambda=\pm(k+1+{p\over n}+{1\over 2n}+\frac{\imu}{2\pi} ),~~~p\in[0,n-1] \cr
{(k+1)(k+2)\over 2} ~~&&\text{for}~~\lambda=\pm(k+3-{p\over n}-{1\over 2n}-\frac{\imu}{2\pi}),~~~p\in[0,n-1] \cr
{(k+1)} ~~&&\text{for}~~\lambda=\pm(k+2-{p\over n}-{1\over 2n}-\frac{\imu}{2\pi} ),~~~p\in[0,p_1]\cr
{(k+1)} ~~&&\text{for}~~\lambda=\pm(k+{p\over n}+{1\over 2n}+\frac{\imu}{2\pi} ),~~~p\in[p_1+1,n-1]
\eea
where $k\in \mathbb{N}$.

The free energy is
\bea
&F_{n}&=\sum_{p=0}^{n-1}\sum_{a=a_1}^{a_2}\left(\zeta'(-2,a)+(3-2a)\zeta'(-1,a)+(a^2-3a+2)\zeta'(0,a)\right)\cr
&+&\sum_{p=0}^{p_1}2(\zeta(-1,c_1)+(1-c_1)\zeta(0,c_1))+\sum_{p=p_1+1}^{n-1}2(\zeta(-1,c_2)+(1-c_2)\zeta(0,c_2))~~~
\eea
with
\be
a_1=1+{p\over q}+{1\over 2q}+\frac{\imu}{2\pi},~~
a_2=3-{p\over q}-{1\over 2q}-\frac{\imu}{2\pi},~~
c_1=2-{p\over q}-{1\over 2q}-\frac{\imu}{2\pi},~~
c_2={p\over q}+{1\over 2q}+\frac{\imu}{2\pi}.
\ee

\begin{figure}[h!]
\includegraphics[width=\textwidth]{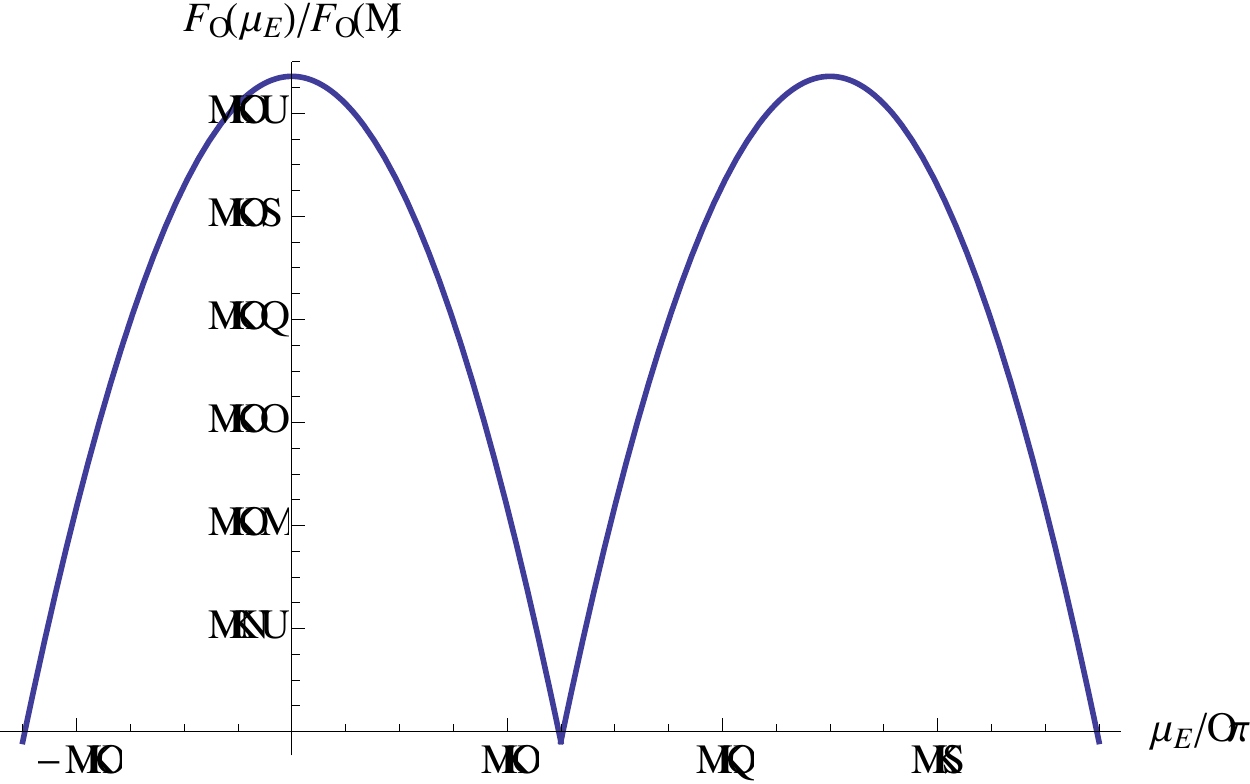}
\caption{Fermion free energy  $n=2$}
\label{free fermion d3 plot}
\end{figure}

The numerical result is shown in Fig.\ref{free fermion d3 plot}.
There are phase transitions at $\frac{\imu}{2\pi}={1\over 2n}+{\mathbb{Z}\over n}$ for the free energy
and $\frac{\imu}{2\pi}={1\over 2n}+{\mathbb{Z}\over n}$ and ${1\over 2}+{\mathbb{Z}}$ for the R\'{e}nyi entropy.

%

%

\section{Holographic computations for $d=2$} \labell{threed}

We now compute the holographic charged \ren entropies for $d=2$ using a
three-dimensional bulk dual.  In this case there are two interesting bulk
duals: Einstein-Maxwell theory and Einstein-Chern-Simons theory.

\subsection{Einstein-Maxwell theory}

Starting with the Einstein-Maxwell action in a three-dimensional bulk
\begin{equation}
I_{E-M} = \frac{1}{2 \lp} \int d^3x \sqrt{-g}\left(\frac{2}{L^2}+\R
- \frac{\lstar^2}{4}F_{\mu\nu}F^{\mu\nu}\right)\,.
\end{equation}
The charged black hole solution analogous to eq.~\reef{bhmetric} for this
theory is \cite{mtz}
\begin{equation}
ds^2 = -f(r) \frac{L^2}{R^2} dt^2 + \frac{dr^2}{f(r)} + r^2 d\theta^2\,,
 \labell{train}
\end{equation}
with
\begin{align}
f(r) &= \frac{r^2}{L^2} - m - \frac{q^2}{2} \log\left(\frac{r}{L\sqrt{m}}\right)\,.
\end{align}
With the above parametrization, the horizon radius is simply $r_H=L\sqrt{m}$.
Note that the geometry is not asymptotically AdS because of the logarithmic
term appearing in $f(r)$. Similar situations were considered in
\cite{Klebanov,Hung,Yau} and this logarithmic behaviour is the signature of
broken conformal invariance, even in the UV. The solution for bulk gauge field
is
 \bea
A=-\frac{qL}{\lstar R}\,\log\!\( {{r\over {L\sqrt{m}}}}\)\, dt \, .
 \labell{gauge2}
 \eea
The integration constant in eq.~\reef{gauge2} was chosen to ensure that $A=0$
at the horizon. However, because of the logarithmic running of the bulk gauge
field, one can not discern the chemical potential and the expectation value of
the dual charge density as easily as in section \ref{Einstein} for $d\ge3$.
Following \cite{Hung,Yau}, we arbitrarily chose a renormalization scale which
will be defined by the radius $r=r_R$. Then the chemical potential is given by
 \beq
  \mue=-{q}\log {{r_{H}\over r_R}}\,.
  \eeq
Hence we find that the chemical potential runs logarithmically with the
renormalization scale.

The temperature of the black hole \reef{train} is
 \beqa
T = \frac{L\, f'(r_H)}{4 \pi R} &=& \frac{r_H}{2\pi R L}\,
\left(1 - \frac{L^2\,\mue^2}{4 r^2_H \log[r_H/r_R]^2}\right) \nonumber\\
&=& T_0\left(x - \frac{\mu^2}{4x \log\left(\frac{xL}{r_R}\right)^2}\right)\,,
\labell{raffle}
 \eeqa
where as before, we have introduced the parameter $x=r_H/L$. The horizon
entropy is given by
\begin{equation}
S = 2\pi r_H = 2\pi L\,x\,.
\end{equation}
The charged \ren entropy for this solution is again determined with
eq.~\reef{ent9}, however, we note that in this integral, both the chemical
potential and the renormalization scale $r_R$ are held fixed. The endpoints of
the integral are again chosen such that $T(x_1,\mue,r_R) = T_0$ and
$T(x_n,\mue,r_R)=T_0/n$. Given the form of eq.~\reef{raffle}, the $x_n$ can
only be solved numerically for given $\mue$ (and $r_R$). Combining the previous
results, we can write the charged \ren entropy as follows:
\begin{equation}
S_n(\mue) = \frac{ \pi L}{n-1} \left( 2 n x_1 - 2 x_n +
n(x_n^2-x_1^2) + \mue (\sqrt{nx_n(nx_n-1)}-n\sqrt{ x_1(x_1-1)})\right)\,.
\end{equation}

%
\begin{figure}[h!]
\includegraphics[width=\textwidth]{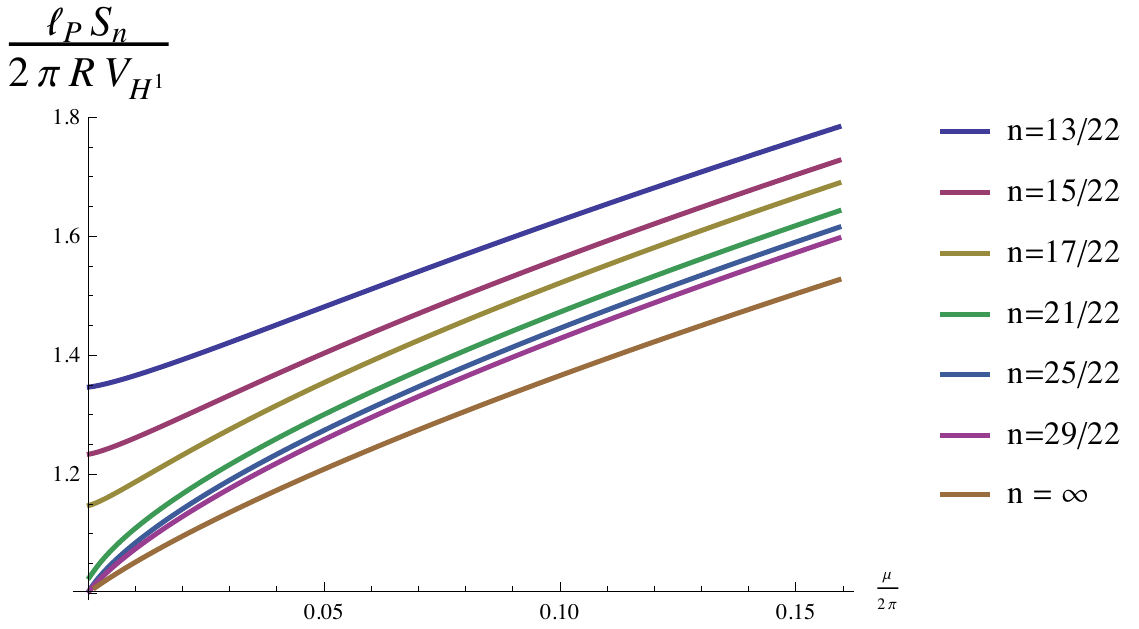}
\caption{Charged \ren entropy evaluated for the charged BTZ black
hole at various values of $n$, setting $L=r_R=1$.}
\label{cbtzplotn}
\end{figure}
\begin{figure}[h!]
\includegraphics[width=\textwidth]{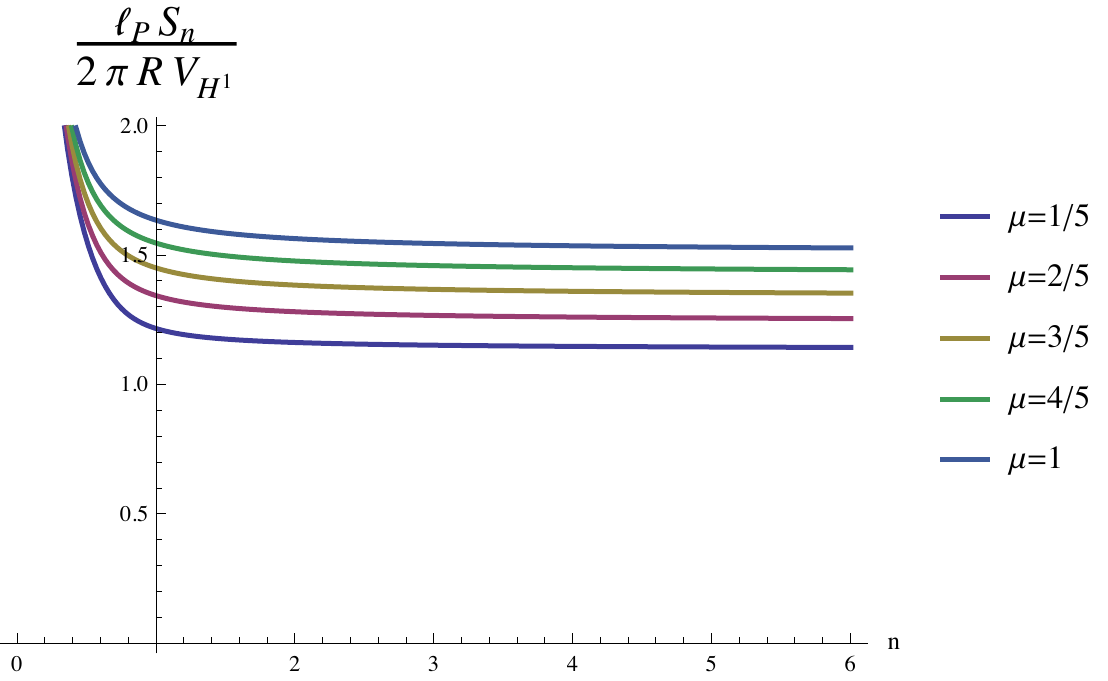}
\caption{Charged \ren entropy evaluated for the charged BTZ black
hole at various values of $\mue$, setting $L=r_R=1$.}
\label{cbtzplot}
\end{figure}
In figure \ref{cbtzplotn}, we show the behaviour of the charge \ren entropy as
a function of $\mu$ for various values of $n$. For large values of $\mu$, all of the
$S_n(\mu)$ appear to increase linearly. 
From figure \ref{cbtzplot}, we see that in the limit $n\to\infty$,
$S_n(\mue)$ seems to approach a finite asymptotic value (which depends on
$\mue$), and we have included a plot in figure \ref{cbtzplotn}. 
This may be contrasted with the behaviour in eq.~\reef{bell} for the same limit
in higher dimensions. Further, the $n\to 0$ limit appears to diverge, in
agreement with the analogous limit in higher dimensions, as given in
eq.~\reef{bell}.

It is again interesting to consider imaginary chemical potentials, which is
accomplished here by replacing $\mue = i \imu$ and $q=i\iq$ in the above
results. The horizon radius for which $T= T_0/n$ is now given by
\begin{equation}
x_n = \frac{1+ \sqrt{1-n^2 \iq^2}}{2 n } \,. \labell{hexagon}
\end{equation}
Clearly, we only have real solutions here when $\iq^2<2/n^2$ and so the charge
can only take values in a finite range. If we re-write eq.~\reef{hexagon} as
 \bea
\left({2n\,x_n}-1\right)^2+n^2\iq^2=1\,,
 \eea
we see the horizon radius and the charge can be parameterized by
 \bea
x_n={1+\cos\phi \over 2 n}\,,\qquad\qquad\iq=\frac{\sin\phi}{n}\,.
 \eea
This is again reminiscent of the free field calculation where analytic
continuation is only possible within a finite window for $\mue$.


\subsection{Einstein-Chern-Simons theory}

We first consider the boundary duals of Einstein-Chern-Simons theory.  The
entropy of the charged BTZ black hole is (see e.g. (6.17) of \cite{Kraus})
 \be
S= 2\pi \left(\sqrt{\frac{c}{6}\left(L_0 - \frac{c}{24} - \frac{1}{4}q^2\right)
}+ \sqrt{\frac{\tilde{c}}{6}\left(\tilde{L}_0 - \frac{\tilde{c}}{24} -
\frac{1}{4}q^2\right) }\right)\,.
 \ee
where $L_0 (\tilde{L}_0)$ and $c(\tilde{c})$ are the Virasoro generator and the
central charge of the left (right) movers. Here $q$ denotes the charge of the
black hole. The expression inside the square root is independent of spectral
flow. In terms of the horizon radius $r_H$,
 \be
L_0 - \frac{c}{24} - \frac{1}{4}q^2 = \frac{\pi\,r_H^2}{2\lp\,L} = \tilde{L}_0
- \frac{\tilde{c}}{24} - \frac{1}{4}q^2\,.
 \ee
The Hawking temperature of the BTZ black hole is
 \be
T= \frac{r_H}{2\pi L^2}\,.
 \ee
Using $c= 12\pi L/\lp$, the entropy is the usual expression
 \be S = 2\pi^2 \frac{L T c}{3} = \frac{\pi c}{3 n}\, . \ee
To obtain the \ren entropy, we integrate $S$. However, since $q$ cancels out,
there is no dependence on the chemical potential. We note that this is in
complete agreement with the free fermion results at least for sufficiently
small $\mu_E$. To be more precise, the above statement is as follows: Since the
gauge potential does not couple to the metric in Chern-Simons theory, the
solution of the equation of motion is a flat connection, $dF=0$, or a constant
gauge potential. Without any source term in the bulk, the gauge potential has
to be zero. The charge $q$ we mentioned above is a charge along the spatial
direction $\theta$. The result suggests that the linear term in the \ren
entropy (\ref{replica-renyi-line}) is not protected by a symmetry or an
anomaly.

\section{Holographic minutiae} \labell{details}

In this appendix, we present various useful details of the holographic
calculations, which are used in section \ref{Einstein}.

\subsection{Energy density}\labell{background}

To evaluate the conformal weight of the twist operators using
eq.~\reef{Tweight}, we need to evaluated the difference of energy densities:
$\E(T_0/n,\mue) - \E(T_0,\mue=0)$. Now in principle, with the introduction of
appropriate boundary counterterms \cite{count1}, one can evaluate each of these
energy densities individually. However, since we only need to determine a
difference of energy densities, it is simpler to use `background subtraction,'
in which case the counterterms play no role.

To begin, we will denote the metric of a surface of constant $r$ by
$\gamma_{\mu\nu} = g_{\mu\nu} - \delta_{\mu r}\delta_{\nu r}/g^{rr}$, and the
boundary hyperbolic metric (\ref{frog}) by $\hat{\gamma}$. Following
\cite{Brown:1992br}, we write the boundary stress tensor
\begin{equation}
\tau_{ab} = \frac{2}{\sqrt{-\gamma}}\frac{\delta I}{\delta \gamma^{ab}}
= \frac{1}{\lp^{d-1}}(\gamma_{ab}K^c_{\phantom{c}c} - K_{ab}),
\end{equation}
where $K_{ab}$ is the extrinsic curvature taken on a regulator surface at some
constant radius $r$. To leading order as $r\rightarrow \infty$,
\begin{equation}
\tau_{00}(T_0/n,\mue)-\tau_{00}(T_0,\mue=0) = \frac{(d-1) L m}{2 \lp^{d-1}R^2 r^{d-2}},
\end{equation}
where $m$ is given by (\ref{mass parameter}). We can then evaluate the energy
density of the boundary field theory with \cite{Myers:1999psa}
\begin{equation}
\begin{split}
&\sqrt{-\hat{\gamma}}\hat{\gamma}^{00}( T_{\tau \tau}(T_0/n,\mue)- T_{\tau \tau}(T_0,\mue=0))\\
 &\ = \lim_{r\rightarrow \infty}\sqrt{-\gamma}\gamma^{00}(\tau_{00}(T_0/n,\mue)-\tau_{00}(T_0,\mue=0)).
\end{split}
\end{equation}
Now using the notation $\E=T_{\tau \tau}$ from the main text, this equation
yields the desired difference of energy densities:
\begin{equation}
\begin{split}
\E(T_0/n,\mue) &- \E(T_0,\mue=0) = \frac{(d-1)L m}{2 \lp^{d-1} R^d}\\
=& \frac{(d-1)}{2 R^d}\,\frac{L^{d-1}}{\lp^{d-1}}  \left(x_n^{\,d-2}(x_n^2-1) + \frac{q^2}{(L\,r_H)^{d-2}}\right)\,.
\end{split}\labell{Edensity}
\end{equation}

\subsection{Charge density} \labell{norm}
In order to determine the magnetic response for our holographic model, we must
evaluate the charge density $\rho(n,\mue) = \langle J_\tau \rangle$ in
eq.~\reef{curweight}. Of course, the standard AdS/CFT dictionary indicates this
expectation value is given by the normalizable component of the gauge field
\reef{gauge}, \ie $\langle J_\tau \rangle\propto q$. However, to make precise
comparisons with the expansion coefficients derived in section \ref{PGtwist},
we need the exact normalization of the current. We evaluate the latter here
with a simple thermodynamic analysis.

Recall that the first law of thermodynamics of our ensemble is given by
 \be
d\E = Tds + \frac{\mu}{2 \pi R} d\rho\,,
 \labell{first1}
 \ee
where $\E$, $s$ and $\rho$ denote the energy, entropy and charge densities
respectively. Hence if the entropy density is held fixed, it follows that
\begin{equation}
\frac{\mue}{2\pi R} = \left(\frac{\partial \E}{\partial \rho}\right)_s
\end{equation}
Now as observed above, we have $\rho(n,\mue) = \alpha\, q$ where $\alpha$ is
some numerical factor which we aim to determine. From eq.~(\ref{thermal}), we
can see that holding the entropy density fixed is equivalent to holding $r_H$
constant. Hence it follows that
\begin{equation}
\begin{split}
\alpha =& \frac{2 \pi R}{\mue} \left(\frac{\partial \E}{\partial q}\right)_{r_H}\\
=& \sqrt{\frac{(d-1)(d-2)}{2}} \frac{\lstar}{\lp^{d-1} R^{d-1}}
\end{split}
\end{equation}
where up to a constant independent of $q$, $\E$ is given by
eq.~(\ref{Edensity}). Therefore, our final result for the charge density is
\begin{equation}
\rho(n,\mue) = \frac{(d-2)x_n^{d-2} }{4\pi\, R^{d-1}}\,\frac{L^{d-3}\lstar^2}{\lp^{d-1}}\,
\mue\,.
 \labell{Qdensity}
\end{equation}

\subsection{Boundary CFT parameters}\labell{two point}

Here, we provide the values of various parameters, \ie $C_V$, $\hc$, $\he$ and
$C_T$, for the boundary CFT dual to the Einstein-Maxwell theory \reef{action}.
This allows us to verify various expressions derived in section \ref{PGtwist}
for the expansion coefficients of the twist operator's conformal weight and
magnetic response within the holographic framework of section \ref{Einstein}.

To begin, we follow the calculation of \cite{witten8} to evaluate the two-point
correlator of a current dual to a bulk Maxwell field, however, we will be
careful to include all of the numerical factors. This allows us to evaluate the
constant $C_V$ appearing in eq.~\reef{twoJ} for the Einstein-Maxwell theory
studied in section \ref{Einstein}. From \cite{witten8}, the bulk solution for
the gauge field $A$, which near the AdS boundary approaches $\lim_{z\to 0} A\to
\sum_i B_i(x) dx^i$, is given by
 \be
A(z,x)= \N \int d^dx'\,\left[ \frac{z^{d-2}}{(z^2+ (x-x')^2)^{d-1}}   B_i(x')
dx^i - z^{d-3} dz \frac{(x-x')^i B_i(x')}{(z^2 + (x-x')^2)^{d-1}}\right]\,,
 \labell{sol}
 \ee
where the normalization constant is given by
 \be
\N = \frac{\Gamma(d-1)}{\pi^{d/2}\Gamma(d/2-1)}\,.
 \labell{norm88}
 \ee
This coefficient $\N$ is chosen to ensure that
 \be
\lim_{z\to 0} \N \frac{z^{d-2}}{(z^2 + x^2)^{d-1}} = \delta^d(x)\,.
 \ee

Now our (Euclidean) Maxwell action \reef{action} is given by
 \be
I_{{Max}}= \frac{\lstar^2}{8 \lp^{d-1}} \int d^{d}xdz \, \sqrt{G}\, G^{AC}
G^{BD}\,F_{AB}F_{CD}\,,
 \labell{Emax}
 \ee
where $A,B$ range over $i=1,\cdots d$ and $z$. Further, we work in Poincar\'e
coordinates where $G_{AB} = (L^2/z^2)\, \delta_{AB}$. To extract the leading
boundary contribution, we only need to consider the terms with $z$ derivatives.
That is,
 \be
I_{Max} =\frac{\lstar^2 L^{d-3}}{4 \lp^{d-1}}\int \frac{dz\,d^dx}{z^{d-3}}\,
\left[(\partial_z A_i)^2  - 2
\partial_i A_z
\partial_z A_i + (\partial_iA_z)^2\right]\,.
 \ee
Using the bulk equations of motion for $A_i, A_z$ and integrating by parts, the
above expression yields a boundary term at $z\to 0$:
 \bea
I_{Max} &&= \lim_{z\to 0}  \frac{\lstar^2 L^{d-3}}{4 \lp^{d-1}} \int
\frac{d^dx}{z^{d-3}} \,\left[
 A_i \partial_z A_i - A_i \partial_i A_z\right]  \nonumber \\
&&= \frac{ (d-1)\N\, \lstar^2 L^{d-3}}{4\lp^{d-1}} \int\int d^dx_1d^dx_2
\frac{B_i(x_1) B_j(x_2)}{|x_{12}|^{2d-2}} \left(\delta_{ij} -
2\frac{x_{12}^ix_{12}^j}{|x_{12}|^2}\right),
 \eea
where the second line follows from substituting in eq.~\reef{sol}. Finally, we
may differentiate this action twice with respect to the sources $B_i(x)$ to
produce the two-point function of the corresponding current and we see the form
matches precisely that given in eq.~\reef{twoJ}. Then we may read off the
central charge $C_V$ in the boundary theory with the normalization of the bulk
Maxwell term given in eq.~\reef{Emax}:
 \be
C_V = \frac{(d-1)\N\,\lstar^2 L^{d-3}}{2 \lp^{d-1}}
=\frac{\Gamma\left(d\right)}{2\pi^{d/2}\,\Gamma\left(d/2-1\right)}\,
\frac{\lstar^2 L^{d-3}}{\lp^{d-1}}\,.
 \labell{cv}
 \ee

Now eq.~\reef{extra} relates this central charge to $(\hc+\he)$, the sum of the
two CFT parameters which define the $\langle TJJ\rangle$ correlator
\reef{3point}. Now for the boundary CFT dual to the Einstein-Maxwell theory
\reef{action}, this correlator does not take the most general form possible
\cite{hofman,subir88}, \ie $\hc$ and $\he$ are not independent parameters.
Rather one finds that
 \be
 \hc=d(d-2)\,\he\,.
 \labell{limit5}
 \ee
Combining this constraint with eq.~\reef{extra}, we can also evaluate $\hc$ and
$\he$ for our holographic theory. In particular, we find
 \be
 \he =
 \frac{\Gamma\left(\frac{d+2}2\right)}{2\pi^{d/2}(d-1)^2}\,C_V
 =\frac{(d-2)\,\Gamma(d+1)}{16\pi^d(d-1)^2}\, \frac{\lstar^2 L^{d-3}}{\lp^{d-1}}
 \labell{housefire}
 \ee
 and then $\hc$ follows from eq.~\reef{limit5}.

Finally, it is convenient to have the central charge $C_T$, which appears in
the two-point correlator of the stress tensor \reef{green}, for our holographic
theory. In this case, the calculation analogous to that above for the Maxwell
field was carried out for the metric in \cite{GBparty}. Hence we can simply
quote the result for $C_T$:
 \be
C_T=\frac{\Gamma(d+2)}{\pi^{d/2}(d-1)\Gamma(d/2)}\,\frac{L^{d-1}}{\lp^{d-1}}\,.
 \labell{CTresult}
 \ee


\begin{thebibliography}{99}

\bibitem 
 {wenx} M.~Levin and X.-G.~Wen,
  ``Detecting Topological Order in a Ground State Wave Function,''
  Phys.\ Rev.\ Lett.\  {\bf 96} (2006) 110405
 [arXiv:cond-mat/0510613];\\
A.~Kitaev and J.~Preskill,
  ``Topological entanglement entropy,''
  Phys.\ Rev.\ Lett.\  {\bf 96} (2006) 110404
  [arXiv:hep-th/0510092];\\
A.~Hamma, R.~Ionicioiu and P.~Zanardi, ``Ground state entanglement and
geometric entropy in the Kitaev's model,'' Phys.\ Lett.\ A {\bf 337} (2005) 22
[arXiv:quant-ph/0406202].

\bibitem 
 {cardy0} P.~Calabrese and J.~L.~Cardy,
  ``Entanglement entropy and quantum field theory,''
  J.\ Stat.\ Mech.\  {\bf 0406}, P002 (2004)
  [arXiv:hep-th/0405152];\\
P.~Calabrese and J.~L.~Cardy, ``Entanglement entropy and quantum field theory:
A non-technical introduction,''
  Int.\ J.\ Quant.\ Inf.\  {\bf 4}, 429 (2006)
  [arXiv:quant-ph/0505193].

\bibitem 
 {rt0}  S.~Ryu and T.~Takayanagi,
  ``Holographic derivation of entanglement entropy from AdS/CFT,''
  Phys.\ Rev.\ Lett.\  {\bf 96} (2006) 181602
  [arXiv:hep-th/0603001];\\
 S.~Ryu and T.~Takayanagi,
  ``Aspects of holographic entanglement entropy,''
  JHEP {\bf 0608} (2006) 045
  [arXiv:hep-th/0605073];\\
T.~Nishioka, S.~Ryu and T.~Takayanagi,
  ``Holographic Entanglement Entropy: An Overview,''
  J.\ Phys.\ A  {\bf 42} (2009) 504008
  [arXiv:0905.0932 [hep-th]];\\
T.~Takayanagi,
  ``Entanglement Entropy from a Holographic Viewpoint,''
  arXiv:1204.2450 [gr-qc].

\bibitem 
 {mvr} M.~Van Raamsdonk,
  ``Comments on quantum gravity and entanglement,''
  arXiv:0907.2939 [hep-th];\\
M.~Van Raamsdonk,
  ``Building up spacetime with quantum entanglement,''
  Gen.\ Rel.\ Grav.\  {\bf 42} (2010) 2323
  [arXiv:1005.3035 [hep-th]].

\bibitem 
 {arch} E.~Bianchi and R.~C.~Myers,
  ``On the Architecture of Spacetime Geometry,''
  arXiv:1212.5183 [hep-th];\\
R.~C.~Myers, R.~Pourhasan and M.~Smolkin,
  ``On Spacetime Entanglement,''
  JHEP {\bf 1306} (2013) 013
  [arXiv:1304.2030 [hep-th]];\\
V.~Balasubramanian, B.~Czech, B.~D.~Chowdhury and J.~de Boer,
  ``The entropy of a hole in spacetime,''
  arXiv:1305.0856 [hep-th].

\bibitem 
 {renyi0} A.~R\'enyi, ``On measures of information and entropy,'' in {\sl Proceedings of the
4th Berkeley Symposium on Mathematics, Statistics and Probability},
{\bf 1}, 547 (U. of California Press, Berkeley, CA, 1961);\\
A.~R\'enyi, ``On the foundations of information theory,'' Rev.\ Int.\ Stat.\
Inst.\ {\bf 33} (1965) 1.

\bibitem 
 {renyi1} For example, see:\\
K.~Zyczkowski, ``Renyi extrapolation of Shannon entropy,'' Open Syst.
Inf. Dyn. {\bf 10}, 297 (2003) [arXiv:quant-ph/0305062];\\
C. Beck and F. Schl\"ogl, ``Thermodynamics of chaotic systems'', (Cambridge
University Press, Cambridge, 1993).

\bibitem 
 {casini9} H.~Casini, M.~Huerta and R.~C.~Myers,
  ``Towards a derivation of holographic entanglement entropy,''
  JHEP {\bf 1105} (2011) 036
  [arXiv:1102.0440 [hep-th]].

\bibitem 
 {renyi} L.-Y.~Hung, R.~C.~Myers, M.~Smolkin and A.~Yale,
  ``Holographic Calculations of Renyi Entropy,''
  JHEP {\bf 1112} (2011) 047
  [arXiv:1110.1084 [hep-th]].

  \bibitem 
 {subir} I.~R.~Klebanov, S.~S.~Pufu, S.~Sachdev and B.~R.~Safdi,
  ``Renyi Entropies for Free Field Theories,''
  JHEP {\bf 1204} (2012) 074
  [arXiv:1111.6290 [hep-th]].

\bibitem 
 {juan} A.~Lewkowycz and J.~Maldacena,
  ``Generalized gravitational entropy,''
  arXiv:1304.4926 [hep-th].

\bibitem 
 {diana} G.~Wong, I.~Klich, L.~A.~P.~Zayas and D.~Vaman,
  ``Entanglement Temperature and Entanglement Entropy of Excited States,''
  arXiv:1305.3291 [hep-th].

\bibitem 
 {gm} P.~Caputa, G.~Mandal and R.~Sinha,
  ``Dynamical entanglement entropy with angular momentum and U(1) charge,''
  arXiv:1306.4974 [hep-th].

\bibitem 
 {susy} T.~Nishioka and I.~Yaakov,
  ``Supersymmetric Renyi Entropy,''
  arXiv:1306.2958 [hep-th].

\bibitem 
 {tweak} H.~Casini,
  ``Entropy inequalities from reflection positivity,''
  J.\ Stat.\ Mech.\  {\bf 1008} (2010) P08019
  [arXiv:1004.4599 [quant-ph]];\\
B.~Swingle, ``Mutual information and the structure of entanglement in quantum
field theory,'' arXiv:1010.4038 [quant-ph].

\bibitem 
 {twistop} L.~Y.~Hung, R.~C.~Myers and M.~Smolkin,
  ``Twist operators in higher dimensions,''
  in preparation.

\bibitem 
 {head} M.~Headrick,
  ``Entanglement Renyi entropies in holographic theories,''
  Phys.\ Rev.\ D {\bf 82} (2010) 126010
  [arXiv:1006.0047 [hep-th]].

\bibitem 
 {Roberge} A.~Roberge and N.~Weiss,
  ``Gauge Theories With Imaginary Chemical Potential and the Phases of {QCD},''
  Nucl.\ Phys.\ B {\bf 275} (1986) 734.

\bibitem 
 {lattice} M.~G.~Alford, A.~Kapustin and F.~Wilczek,
  ``Imaginary chemical potential and finite fermion density on the lattice,''
  Phys.\ Rev.\ D {\bf 59} (1999) 054502
  [hep-lat/9807039].


\bibitem 
 {Witten-Index} E.~Witten,
  ``Constraints on Supersymmetry Breaking,''
  Nucl.\ Phys.\ B {\bf 202} (1982) 253.

\bibitem 
 {cthem} R.~C.~Myers and A.~Sinha,
  ``Seeing a c-theorem with holography,''
  Phys.\ Rev.\ D {\bf 82} (2010) 046006
   [arXiv:1006.1263 [hep-th]];\\
  R.~C.~Myers and A.~Sinha,
  ``Holographic c-theorems in arbitrary dimensions,''
JHEP {\bf 1101} (2011)  125 [arXiv:1011.5819 [hep-th]].

\bibitem 
 {Eric} E.~Perlmutter,
  ``A universal feature of CFT Renyi entropy,''
  arXiv:1308.1083 [hep-th].

\bibitem 
 {PetkouOsborn} H.~Osborn and A.~C.~Petkou,
``Implications of conformal invariance in field theories for general
dimensions,'' Annals Phys.\  {\bf 231} (1994) 311
  [hep-th/9307010].

\bibitem 
 {tadashi-twist} T.~Takayanagi, unpublished.

\bibitem 
 {ant}  H.~Casini and M.~Huerta,
  ``Entanglement entropy in free quantum field theory,''
  J.\ Phys.\ A {\bf 42} (2009) 504007
  [arXiv:0905.2562 [hep-th]];\\
  H.~Casini, C.~D.~Fosco and M.~Huerta,
  ``Entanglement and alpha entropies for a massive Dirac field in two dimensions,"
  J.Stat.Mech. 0507 (2005) P07007
   [arXiv:cond-mat/0505563];\\
 T.~Azeyanagi, T.~Nishioka and T.~Takayanagi,
  ``Near Extremal Black Hole Entropy as Entanglement Entropy via AdS(2)/CFT(1),''
  Phys.\ Rev.\ D {\bf 77} (2008) 064005
  [arXiv:0710.2956 [hep-th]].

\bibitem 
 {renyiphases} A.~Belin, A.~Maloney and S.~Matsuura,
  ``Holographic Phases of Renyi Entropies,''
  arXiv:1306.2640 [hep-th].

\bibitem 
 {thermodynamics} R.-G.~Cai and A.~Wang,
  ``Thermodynamics and stability of hyperbolic charged black holes,''
  Phys.\ Rev.\ D {\bf 70} (2004) 064013
  [hep-th/0406057].

\bibitem 
 {holographicsuperconductor} S.~A.~Hartnoll, C.~P.~Herzog and G.~T.~Horowitz,
  ``Holographic Superconductors,''
  JHEP {\bf 0812} (2008) 015
  [arXiv:0810.1563 [hep-th]].

\bibitem{Belin:2014mva} 
  A.~Belin, L.~Y.~Hung, A.~Maloney and S.~Matsuura,
  JHEP {\bf 1501}, 059 (2015)
  [arXiv:1407.5630 [hep-th]].

\bibitem 
 {GBBH1} M.~Cvetic, S.~Nojiri and S.~D.~Odintsov, ``Black
    hole thermodynamics and negative entropy in de Sitter and anti-de Sitter
    Einstein-Gauss-Bonnet gravity,''
  Nucl.\ Phys.\ B {\bf 628} (2002) 295
  [hep-th/0112045].

\bibitem 
 {GBBH2} X.-H.~Ge, Y.~Matsuo, F.-W.~Shu, S.-J.~Sin and T.~Tsukioka,
``Viscosity Bound, Causality Violation and Instability with Stringy Correction
and Charge,''
  JHEP {\bf 0810} (2008) 009
  [arXiv:0808.2354 [hep-th]].

\bibitem 
 {anninos} D.~Anninos and G.~Pastras,
``Thermodynamics of the Maxwell-Gauss-Bonnet anti-de Sitter Black Hole with
Higher Derivative Gauge Corrections,''
  JHEP {\bf 0907} (2009) 030
  [arXiv:0807.3478 [hep-th]].

\bibitem 
 {spin2} A.~Belin, L.-Y.~Hung, A.~Maloney, S.~Matsuura, R.C.~Myers
and T.~Sierens, in preparation.

\bibitem 
 {heatkernel} A.~Grigor'yan and M. Noguchi,
 ``The heat kernel on hyperbolic space, ''
 Bull.\ London Math.\ Soc.\  {\bf 30} (1998) 643;\\
 A.~Grigor'yan,
 ``Upper bounds on a complete non compact manifold, ''
 J.\ Funct.\ Anal.\  {\bf 127} (1995) 363;\\
 A.~Debiard,~B.~Gaveau,~E.~Mazet,
 `` Theoreme de comparison in geometrie riemannienne,''
 Publ.\ Res.\ Inst.\ Math.\ Sci.\ Kyoto {\bf 12} (1976) 391;\\
   R.~Camporesi and A.~Higuchi,
  ``Spectral functions and zeta functions in hyperbolic spaces,''
  J.\ Math.\ Phys.\  {\bf 35} (1994) 4217;\\
 R. Camporesi,
 Commun. Math. Phys. 148 (1992) 283;\\
   A.~Lewkowycz, R.~C.~Myers and M.~Smolkin,
  ``Observations on entanglement entropy in massive QFT's,''
  JHEP {\bf 1304} (2013) 017
  [arXiv:1210.6858 [hep-th]].

\bibitem 
 {density-of-states} R.~Camporesi,
  ``Harmonic analysis and propagators on homogeneous spaces,''
  Phys.\ Rept.\  {\bf 196} (1990) 1;\\
  R.~Camporesi,
  ``The Spinor heat kernel in maximally symmetric spaces,''
  Commun.\ Math.\ Phys.\  {\bf 148} (1992) 283;\\
  A.~A.~Bytsenko, G.~Cognola, L.~Vanzo and S.~Zerbini,
  ``Quantum fields and extended objects in space-times with constant curvature spatial section,''
  Phys.\ Rept.\  {\bf 266} (1996) 1
  [hep-th/9505061].

\bibitem 
 {mtz} C.~Martinez, C.~Teitelboim and J.~Zanelli,
  ``Charged rotating black hole in three space-time dimensions,''
  Phys.\ Rev.\ D {\bf 61} (2000) 104013
  [hep-th/9912259].


\bibitem 
 {Klebanov} I.~R.~Klebanov and E.~Witten,
  ``AdS/CFT correspondence and symmetry breaking,''
  Nucl.\ Phys.\ B {\bf 556} (1999) 89
  [hep-th/9905104].

\bibitem 
 {Hung} L.-Y.~Hung and A.~Sinha,
  ``Holographic quantum liquids in 1+1 dimensions,''
  JHEP {\bf 1001} (2010) 114
  [arXiv:0909.3526 [hep-th]].

\bibitem 
 {Yau} A.~Ejaz, H.~Gohar, H.~Lin, K.~Saifullah and S.-T.~Yau,
  ``Quantum tunneling from three-dimensional black holes,''
  arXiv:1306.6380 [hep-th].

\bibitem 
 {Kraus} P.~Kraus,
  ``Lectures on black holes and the AdS(3)/CFT(2) correspondence,''
  Lect.\ Notes Phys.\  {\bf 755} (2008) 193
  [hep-th/0609074].

\bibitem 
 {count1} V.~Balasubramanian and P.~Kraus,
  ``A Stress tensor for Anti-de Sitter gravity,''
  Commun.\ Math.\ Phys.\  {\bf 208} (1999) 413
  [hep-th/9902121];\\
R.~Emparan, C.~V.~Johnson and R.~C.~Myers,
  ``Surface terms as counterterms in the AdS/CFT correspondence,''
  Phys.\ Rev.\ D {\bf 60} (1999) 104001
  [hep-th/9903238].

  \bibitem 
  {Brown:1992br} J.~D.~Brown and J.~W.~York, Jr.,
    ``Quasilocal energy and conserved charges derived from the gravitational action,''
    Phys.\ Rev.\ D {\bf 47}, 1407 (1993)
    [gr-qc/9209012].

    \bibitem 
     {Myers:1999psa} R.~C.~Myers,
      ``Stress tensors and Casimir energies in the AdS / CFT correspondence,''
      Phys.\ Rev.\ D {\bf 60} (1999)  046002
      [hep-th/9903203].

\bibitem 
 {witten8} E.~Witten,
  ``Anti-de Sitter space and holography,''
  Adv.\ Theor.\ Math.\ Phys.\  {\bf 2} (1998) 253
  [hep-th/9802150].

\bibitem 
 {hofman} D.~M.~Hofman and J.~Maldacena,
  ``Conformal collider physics: Energy and charge correlations,''
  JHEP {\bf 0805} (2008) 012
  [arXiv:0803.1467 [hep-th]].

\bibitem 
 {subir88} D.~Chowdhury, S.~Raju, S.~Sachdev, A.~Singh and P.~Strack,
  ``Multipoint correlators of conformal field theories: implications for quantum critical transport,''
  Phys.\ Rev.\ B {\bf 87} (2013) 085138
  [arXiv:1210.5247 [cond-mat.str-el]].



\bibitem 
 {GBparty} A.~Buchel, J.~Escobedo, R.~C.~Myers, M.~F.~Paulos, A.~Sinha and M.~Smolkin,
  ``Holographic GB gravity in arbitrary dimensions,''
  JHEP {\bf 1003} (2010) 111
  [arXiv:0911.4257 [hep-th]].



\end{thebibliography}
\end{document}